# TRADING OFF ENERGY STORAGE AND PAYLOAD – AN ANALYTICAL MODEL FOR FREIGHT TRAIN CONFIGURATION

**Max T.M. Ng**
Graduate Research Assistant
Transportation Center
Northwestern University
600 Foster Street
Evanston, IL 60208, USA
Email: maxng@u.northwestern.edu

**Adrian Hernandez**
Graduate Research Assistant
Department of Civil and Environmental Engineering
Northwestern University
2145 Sheridan Road
Evanston, IL 60208, USA
Email: adrianhernandez2025@u.northwestern.edu

**Pablo L. Durango-Cohen**
Associate Professor
Department of Civil and Environmental Engineering
Northwestern University
2145 Sheridan Road
Evanston, IL 60208, USA
Email: pdc@northwestern.edu

**Hani S. Mahmassani**
William A. Patterson Distinguished Chair in Transportation
Director, Transportation Center
Northwestern University
600 Foster Street
Evanston, IL 60208, USA
Email: masmah@northwestern.edu

**ABSTRACT**

To support planning of alternative fuel technology (e.g., battery-electric locomotives) deployment for decarbonizing non-electrified freight rail, we develop a convex optimization formulation with a closed-form solution to determine the optimal number of energy storage tender cars in a train. The formulation shares a similar structure to an Economic Order Quantity (EOQ) model. For given market characteristics, cost forecasts, and technology parameters, our model captures the trade-offs between inventory carrying costs associated with trip times (including delays due to charging/refueling) and ordering costs associated with train dispatch and operation (energy, amortized equipment, and labor costs).

To illustrate the framework, we find the optimal number of battery-electric energy tender cars in 22,501 freight markets (origin-destination pairs and commodities) for U.S. Class I railroads. The results display heterogeneity in optimal configurations with lighter, yet more time-sensitive shipments (e.g., intermodal) utilizing more battery tender cars. For heavier commodities (e.g., coal) with lower holding costs, single battery tender car configurations are generally optimal. The results also show that the optimal train configurations are sensitive to delays associated with recharging or swapping tender cars.

**Keywords:**

# 1    INTRODUCTION

For over a century, freight rail has been the backbone of the American economy, accounting for around 40% of long-haul freight movements by annual ton-miles, while contributing less than 2% of the greenhouse gas emissions of the transportation sector (Association of American Railroads, 2022a). Despite the efficiency relative to other modes such as trucking, major railroads have committed to voluntary carbon emissions reductions (Association of American Railroads, 2022b) without electrifying the tracks (Association of American Railroads, 2021a). Synergistic with the deployment of electric vehicles, research and development of alternative fuel technologies such as battery-electric locomotives have gained traction in recent years, as railroads, locomotive manufacturers, and start-ups entered the market, and are conducting (localized) test operations of these new technologies in North America (BNSF Railway, 2021; de Chant, 2022; Johnson, 2022; Tangermann, 2022; Union Pacific, 2022) and other parts of the world such as Australia, Brazil, and France (Molitor, 2022; Vantuono, 2023; Wabtec Corporation, 2023).

Until recently, there had been a dearth of literature on the development of deployment strategies for the rapidly evolving energy technologies on freight trains. In response to the urgency of climate change, the U.S. Department of Energy Advanced Research Project Agency-Energy (ARPA-E) has funded a number of research initiatives to analyze the economic, environmental, and operational impacts, including LOCOMOTIVES (Ahuja et al., 2023; Aredah et al., 2024; Baker et al., 2023; Hernandez et al., 2024) and INTERMODAL (ARPA-E, U.S. Department of Energy, 2023). These efforts are echoed by agencies across the world, such as Europe's Rail Joint Undertaking (EU-Rail) in Europe (Chamaret et al., 2023), National Natural Science Foundation of China (NSFC) and National Railway Administration in China (National Railway Administration, 2024; L. Wang et al., 2021), and East Japan Railway Company (JR-East) in Japan (East Japan Railway Company, 2022; Kadono, 2023). While battery-electric technology is identified as promising zero emissions technologies (International Energy Agency, 2019) with their environmental benefits demonstrated (Cipek et al., 2021), its limitations in range and trade-offs with revenue tonnage restrict emissions reduction potentials (Fullerton and Dick, 2015). Technological solutions, such as solid state batteries may improve the capabilities of advanced fuel propulsion systems; however, significant uncertainties surround their availability and economic viability, motivating the need to explore alternative strategies to deploy these technologies in order to attain their full potential benefits (Wanitschke and Hoffmann, 2020).

Advanced fuel propulsion systems relying on batteries or hydrogen fuel cells impose constraints on locomotive range compared to diesel, due to their lower volumetric energy densities and train limitations in onboard energy storage capacity. To address this issue, recent work discusses energy storage tender car attachments to extend locomotive ranges (Valentine, 2021). Energy storage tender cars—rail cars with energy storage media to power connected locomotives, also known as On-Board Energy Storage Systems, are not a new concept for bypassing space constraints onboard locomotives to extend train ranges (Iden, 2014; Simpson, 2018). Popovich et al. (2021), for example, discuss the technical feasibility of electricity transmission from tender cars to locomotives via cables. Their economic and environmental assessments of a 14-MWh battery configuration with a (relatively modest) range of 241 km show reductions in greenhouse gas emissions and financial competitiveness with diesel-electric operations for intermodal trains with 1700-ton payloads, however without considering operational impacts, such as delay.

The motivation for our work is that, even after selecting a fuel propulsion system for a given freight market, different train configurations lead to different capabilities and costs. Specifically, the number of energy storage tender cars determines a train's payload, i.e., the number of revenue cars. The rate at which energy storage tender cars displace revenue cars is a function of technical elements such as freight characteristics, locomotive power, topography, travel speed, and track geometry and condition. For example, Figure 1 illustrates a situation where one energy storage tender car replaces two revenue cars. In turn, payload determines the number of trains/trips required to satisfy demand, and therefore, the associated capital, labor, and fuel costs. At the same time, on-board energy storage reduces the number of required service/refueling/swapping stops, thereby decreasing trip duration. Thus, the number of tender cars also determines inventory carrying costs (delay costs due to charging/refueling). Together, these observations constitute the motivation and account for the contribution of the work herein.

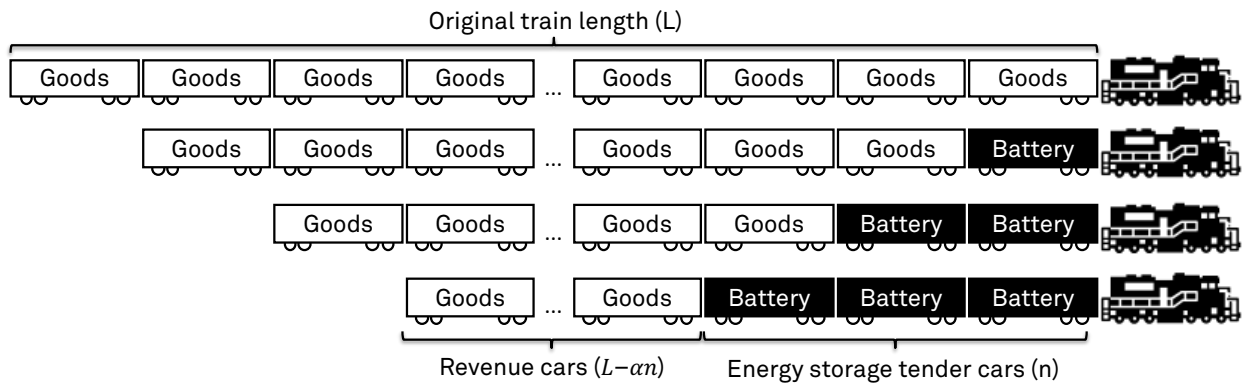

**Figure 1 – Illustration of Trade-off between the Payload and Energy Storage Tender Cars**

We formulate a convex optimization model that captures the fundamental trade-offs between payload and range to determine the optimal train configuration for a given freight market. Conceptually, the model shares a similar structure to an economic order quantity (EOQ) model, where the order quantity determines the order frequency and the (average) inventory held (per period), as well as the associated ordering, inventory-carrying, and purchasing costs (Harris, 1913).[1] It provides a closed-form solution of the optimal number of energy storage tender cars based on market characteristics, cost forecasts, and technology parameters.

Under a set of technological assumptions on future battery-electric trains on non-electrified tracks, we demonstrate the applicability of the model with numerical examples in three real-world freight markets. We then extend the demonstration to 22,501 freight markets of different railroads, origin-destination pairs, and commodities in the Carload Waybill Sample (Surface Transportation Board, 2019a) for U.S. Class I railroad traffic in 2019 for a macroscopic analysis of battery-electric technology deployments for different regions, shipment commodities, delay costs, charging speeds, and capital costs. The goal of this application is to investigate whether configurations of multiple battery tender car assignments produce cost savings. To this end, we assess the optimal tender configurations with respect to commodity-specific characteristics such as value of time and energy requirements, to determine which scenarios favor more or fewer battery tender cars. This framework enables railroads and authorities to assess the benefits of energy storage tenders, thereby preparing and planning for the roll-out of new energy technologies to support rail decarbonization.

---

[1] In a basic EOQ model, purchasing costs are not functions of the order quantity.

The remainder of the paper is organized as follows. Related work is reviewed in Section 1 with an emphasis on benchmarking the proposed model to determine train configurations with frameworks appearing in synergistic settings, e.g., sizing shipments in trucking or freight train applications. The model is presented in Section 2. Results from linehaul examples for intermodal, automotive, and coal shipments are presented in Section 3, followed by an analysis of the network-wide results in Section 4. Section 5 concludes the paper with a discussion of contributions and implications of the findings, along with suggested future research directions.

## 1   LITERATURE REVIEW

The literature in rail decarbonization is rapidly growing along with the urgency of climate actions (see recent reviews by Ahsan et al. (2023), freight specifically by Gandhi et al. (2022)). Many research, including Zenith et al. (2020), Aredah et al. (2024), Hernandez et al. (2024), and others listed in Table 1 compare the costs and benefits of different freight rail energy technologies (e.g., battery-electric, hydrogen, biofuel, e-fuel). While energy storage systems are increasingly deployed in electrified railway systems (Liu and Li, 2020), on-board energy storage offers a green alternative to some other freight rail systems, such as in Australia and the U.S. where track electrification is less common due to economic or technical considerations (Association of American Railroads, 2021a; Knibbe et al., 2022). Popovich et al. (2021) demonstrate the financial viability and environmental benefits of battery-electric energy with tender car attachments. They, together with Moraski et al. (2023), also assess the impacts of battery-electric trains on the electric grid and suggest such energy storage can be used to enhance grid resilience.

**Table 1 - Summary of Selected Recent Literature on Freight Rail Decarbonization**

| Literature | Alternative Fuel Technology | Cost Evaluation | Sizing of Energy Storage | Region |
|---|---|---|---|---|
| Aredah et al. (2024) | Battery-electric, electrification, biofuel, e-Fuel, hydrogen | Emissions, energy | Pre-set | U.S. railroads |
| Fullerton and Dick (2015) | Electrification, hybrid diesel-battery-electric, LNG | Delay, emissions, energy | Pre-set | N/A *(theoretical case)* |
| Hernandez et al. (2024) | Battery-electric, biofuel, e-Fuel, hydrogen | Battery, delay, emissions, energy, locomotive | Pre-set | U.S. Class I railroads |
| Knibbe et al. (2022) | Battery-electric, hydrogen | Battery, emissions, energy, locomotive | Based on route requirements | Several lines in Australia |
| Knibbe et al. (2023) | Hybrid hydrogen-battery | Battery, emissions, energy, locomotive | Based on route requirements | Several lines in Australia |

| Popovich et al. (2021) | Battery-electric | Battery, emissions, energy, locomotive | Based on route requirements | One line in California, U.S. |
|---|---|---|---|---|
| Zenith et al. (2020) | Battery-electric, electrification, hybrid hydrogen-battery, hydrogen | Battery, emissions, energy, locomotive | Based on route requirements | Two lines in Norway and U.S. |
| This work | Energy storage technology (case study in battery-electric) | Battery, delay, emissions, energy, locomotive | Optimized based on economic costs | 22,541 routes of U.S. Class I railroads |

The concept of adding battery tender cars to extend ranges of battery-electric locomotives has been discussed in recent literature (Iden, 2014; Simpson, 2018), with their mechanical and electric specifics introduced by Iden (2021) and Barbosa (2023). Knibbe et al. (2022) delineate the use cases of battery tenders and hydrogen-fuel-cell tenders to address range limitations for Australian rail freight. Its extension (Knibbe et al., 2023) further optimizes the tender sizes, albeit without considering the operational impacts. While the benefits of adopting energy storage tender cars have been evaluated by Popovich et al. (2021) and Knibbe et al. (2022), they do not include an important operational consideration—delays (Fullerton and Dick, 2015). As summarized in Table 1, this paper addresses the research gap in optimizing the consequential freight train configuration, specifically the number of energy storage tender cars relative to the payload per train under the consideration of operational delays.

This optimization problem bears resemblance to problems seen in other modes of transportation. For aircraft, the payload-range diagram illustrates the trade-offs brought by constraints in aerodynamics as well as fuel and passenger/cargo capacities (Ackert, 2013). To achieve longer ranges, the payload is reduced in exchange for more fuel or less total weight in the case of full fuel tanks. Chao and Hsu (2014) show that the optimal payload varies with routes, aircraft types, and fuel costs for air cargo. In freight rail, the energy storage tender car problem corresponds to the trade-off between fuel and payload, as an additional energy storage tender car means extra weight must be hauled by the locomotives. A similar problem may be delay costs due to shipment consolidation (less frequent trains) in exchange for economies of scale (Keaton, 1991). This paper considers the delay costs due to less energy storage (and higher payload) in exchange for fewer locomotives/trains trips in a linehaul market.

To determine optimal shipment sizes based on costs of transport, inventory, and capital investments, EOQ models are a core component of inventory management theory (Burns et al., 1985). They are widely adopted in the transportation and logistics field with several extensions reviewed by Silver (1981) and Combes (2014). Generally, optimal shipment sizes increase with haulage distance and weight as shown in the cost derivation by McCann (2001). These models have also been demonstrated empirically, e.g., with country-wide shipments in France (Combes, 2012) and intra-city shipments in Japan (Sakai et al., 2020). The optimal delivery lot sizes from EOQ models can be used as a microscopic decision output to inform more aggregate production quantities of firms (Zhao et al., 2015) and combined with vehicle size choice (Abate and de Jong, 2014). EOQ models have also been extended for other elements including mode choices (Chen, Roberts & Ben-Akiva, 1978; Winston 1979), as recently reviewed by Engebrethsen and Dauzère-Pérès (2019). Despite the application of multiple EOQ model variants to various problem types, they have not been applied to study interactions between vehicle range and payload. Energy storage tenders introduce a payload problem where range extensions—to cut delay costs—reduce train payloads, potentially requiring additional train dispatchment. These features lead to the model form presented in Section 2.

## 2 FORMULATION

In this section, we present the formulation of the convex optimization problem for determining the optimal number of energy storage tender cars with a closed-form solution. We assume that a given market's characteristics, i.e., the type, mix, and quantities of goods that are shipped, the trip distance, topography, track characteristics, level-of-service requirements, etc., determine operating parameters including the length, $L$, i.e., total number of cars, per train. Together, market characteristics, operating parameters, and other technology-specific attributes, determine a train's range, which, for simplicity, we assume to be a linear function of the number of energy storage tender cars, $n$, on a train, i.e., $R(n) = r \cdot n$, where $r$ is the range (km) provided by the energy stored in one tender car. It follows that the required number of refueling/recharging/swapping stops for a trip distance $D$ is $s(n) = D/R(n)$.[2] The optimal number of energy storage tender cars per train, $n^*$, minimizes the cost to ship the total demand, $Q$ (cars/year). The total costs per year, $TC$ (\$/year), are associated with equipment, i.e., locomotives and batteries, use,[3] charging/fueling, and inventory carrying costs, as given in (1) below:

$$TC = k \cdot \frac{Q}{L - \alpha n} + h \cdot t(n) \cdot Q + p \cdot Q \qquad (1)$$

The first term represents the fixed costs associated with train dispatch. These costs are analogous to the order costs in an EOQ model, and in the case of trains are associated with equipment, fuel, and labor. $L - \alpha n$ is the payload (cars) per train and $\alpha$ is the ratio of the weight of an energy storage tender car to that of an average loaded railcar. Thus, $Q/(L - \alpha n)$ corresponds to the number of trains required to satisfy the annual demand. The fixed costs per train are denoted by $k$ (\$/train) and are assumed to be constant with respect to the number of energy storage cars, as in the standard form EOQ. These constant fixed costs represent distance-based fixed costs of equipment operation, fuel, and labor.[4]

The second term corresponds to the inventory-carrying costs applied to in-transit cars, where $h$ is the holding cost per unit car per hour. The payload, $L - \alpha n$, is held for the entire duration of the trip, $t(n)$ (hours), and thus, the time-weighted average number of in-transit cars per year is given by $(L - \alpha n) \cdot t(n) \cdot \frac{Q}{L - \alpha n}$, which explains the above expression.

---

[2] The exact number of refueling/recharging/swapping stops required is $s(n) = \lceil D/R(n) \rceil - 1$ when the tender cars are fully refueled/recharged at the beginning, i.e., the refueling/recharging time taken at origin/destination is ignored. However, the formulation is shown with $s(n) = D/R(n)$ for simplicity and without loss of generality.

[3] We assume that market characteristics determine the number of locomotives, i.e., power requirements, and the length, i.e., total number of cars, per train.

[4] The fixed costs can also be formulated as a function of the number of energy storage tender cars for the case where individual capital costs for locomotives and energy storage tenders also depend on the journey time. Though this case leads to a more complex formulation, the mathematical results in Appendix A show it maintains the desired qualities of the standard model.

The last term in (1) is analogous to the purchasing costs in an EOQ model; $p$ represents the variable cost of transporting a rail car. Because these costs are not functions of $n$, without loss of generality, we assume $p = 0$.

The function for the trip time, $t(n)$, is specified in (2) as follows. The first term, $t_0$, represents the nominal trip duration and the second term represents the total time to refuel/recharge/swap tender cars, where $t_s$ is the time per stop.

$$t(n) = t_0 + s(n) \cdot t_s \tag{2}$$

The fixed costs per train, $k$, can be considered constant, as in (3) as follows.[5] The first component, $c_l$, captures the costs per train of locomotive equipment and operation, while $c_n$ represents the costs per train of energy tender car equipment and operation. The final term uses $f$, the energy cost per energy tender car per stop, and reduces to $f\,D/r$, which is invariant with respect to $n$.

$$k(n) = c_l + c_n + f \cdot n \cdot s(n) \implies k = c_l + c_n + f\frac{D}{r} \tag{3}$$

For model realism, we make the following assumptions on train composition: $n \geq 1$, as we require a non-zero energy source (i.e., at least one energy storage tender car), $L - \alpha n \geq 1$, as we must be moving non-zero carloads in addition to the energy tender cars, and $L \geq 2$, at minimum, the train consists of one energy tender car and one revenue car. These assumptions bound $n \in [1, \frac{L-1}{\alpha}]$ and $L \in [2, \infty)$. Figure 2 illustrates the variations of these costs with the number of energy storage tender cars, $n$, broken down into the fixed (order) cost and holding cost components.

---

[5] The assumption of constant fixed cost per train is closer to realism when the capital costs, in particular the capital costs of energy storage tenders, are negligible. The full formulation in Appendix A considering time-based capital costs illustrates results with similar behavior.

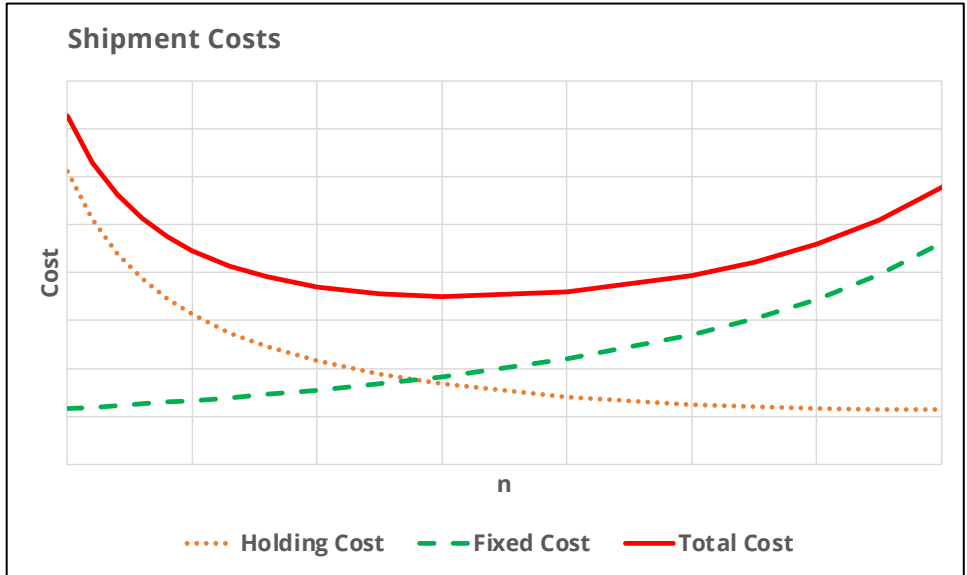

**Figure 2 – Illustration of Holding and Fixed (Order) Costs varying with the Number of Energy Storage Tender Cars, $n$**

The first derivative of the total cost function in (4) shows the order costs are monotonically increasing in $n$, whereas the holding costs are monotonically decreasing in $n$, over the defined ranges for $n$ and $L$, verifying the trade-off between these two components seen in Figure 2. Furthermore, the total cost function is proven to be convex, as its second derivative in (5) is positive. This guarantees that an optimal solution, $n^*$, exists as a global minimum, as in (6).

$$TC' = kQ \frac{\alpha}{(L - \alpha n)^2} - h \frac{D}{r} t_s Q \frac{1}{n^2} \tag{4}$$

$$TC'' = kQ \frac{2\alpha^2}{(L - \alpha n)^3} + 2h \frac{D}{r} t_s Q \frac{1}{n^3} > 0 \tag{5}$$

$$n^* = \frac{L}{\alpha + \sqrt{\frac{k\alpha}{h t_s} \frac{r}{D}}} \tag{6}$$

$$R^* = r \cdot \frac{L}{\alpha + \sqrt{\frac{k\alpha}{h t_s} \frac{r}{D}}} \tag{7}$$

$$TC^* = \frac{Q}{L} \cdot \left[ k + \alpha h \frac{D}{r} t_s + 2 \sqrt{k \alpha h \frac{D}{r} t_s} \right] + h t_0 Q \tag{8}$$

The optimal number of energy storage tender cars per train in (6) leads to an economically optimal train range, as in (7) and a minimum total cost in (8). In the results for the optimal number of energy storage tender cars in (6), we observe the balancing of order and holding costs in this model. Furthermore, due to the convexity of the total cost function, to obtain an optimal integer number of energy storage tender cars, it is sufficient to check the two adjacent integer values.

The application of the model with the full formulation in Appendix A, considering time-dependent fixed costs and energy storage tender capital costs, is demonstrated for the case of future battery-electric energy storage technology, with three numerical examples of linehaul rail freight in Section 3 and network-wide freight market analysis in Section 4.

## 3    NUMERICAL EXAMPLES – LINEHAUL SHIPMENTS

The three selected linehaul examples represent different commodities and distances in the Western American railroads (Table 2). The first two are relatively lighter shipments of intermodal and automotive goods between Los Angeles and Chicago, each with different delay costs. The last example is a heavier, yet slower coal freight movement between Powder River Basin, Wyoming and Chicago. Results for the number of battery tender cars are normalized per locomotive, as $n/n_l$, for easy comparison, as the exact number of batteries would vary with the train length and the number of locomotives.

**Table 2 – Characteristics of Linehaul Examples**

| Parameter | Intermodal Freight between Los Angeles and Chicago | Automotive Freight between Los Angeles and Chicago | Coal Freight between Powder River Basin and Chicago |
|---|---|---|---|
| Trip distance (miles), $D$ (mile) | 2,300 | 2,300 | 1,400 |
| Nominal trip duration, $t_0$ (hours) | 75.7 | 109 | 70.7 |
| Train length, $L$ (car) | 118 | 118 | 73 |
| Range provided by the energy stored in one tender car, $r$ (mile) | 62 | 76 | 320 |
| Total demand, $Q$ (car/year) | 1,500 | 3,000 | 1,000 |
| Ratio of the weight of a battery tender car to that of an average loaded railcar, $\alpha$ | 10.4 | 10.4 | 1.3 |
| Holding cost, $h$ ($/car-hour) | 32 | 9.5 | 9.5 |
| Time per stop, $t_s$ (hour) | 3.73 | | |
| Refueling cost per energy tender car per stop, $f$ ($) | 2,240 | | |

Cost evaluation is based on multiple data sources of rail operations and battery-electric technologies. All costs are shown in 2019 U.S. Dollar values. Rail data, including train speed, average number of cars per train, and locomotive utilization, are collected from Class I railroad submissions (Surface Transportation Board, 2019b, 2019c). Technical parameters including battery weight and capacity are referenced from (Popovich et al., 2021). The underlying parameters and assumptions are discussed in detail in Appendix B.

The detailed results are tabulated in Appendix C.

## 3.1   Intermodal Freight between Los Angeles and Chicago

Relative to other goods, intermodal freight is generally lighter but is shipped on faster trains, which accounts for the high energy requirements per ton-mile (see Table B.3 in Appendix B). The results shown in Figure 3 suggest an optimum is attained at four batteries per locomotive, with a 248-mile range and nine stops over the 2,300-mile route. There is an obvious trade-off between the holding costs (delay) and the order costs (in particular the fixed battery and charging costs), while the increase in the fixed locomotive costs is less impactful.

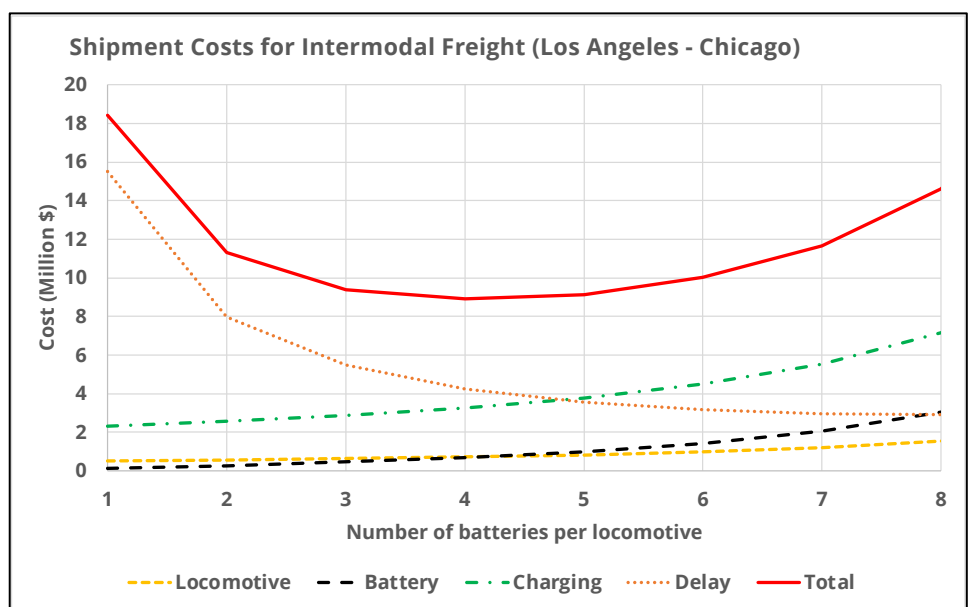

**Figure 3 – Costs of Shipments for Intermodal Freight between Los Angeles and Chicago**

This example illustrates a time-sensitive shipment (reflected in the high unit delay cost). It is worth exploring whether the optimal battery-electric configuration demonstrated here is competitive with other alternative energy sources in terms of operational impacts, as it involves a 37-hour delay time, which may be considered unacceptable in many markets, or in comparison with trucking.

## 3.2   Automotive Goods between Los Angeles and Chicago

With the same origin-destination pair and similar energy intensity, this example showcases an automotive goods movement with a lower optimal number of batteries (higher payload) due to the lower delay costs (Figure 4). The total cost is the lowest at three batteries per locomotive, with a 229-mile range and ten stops per 1,000 miles.

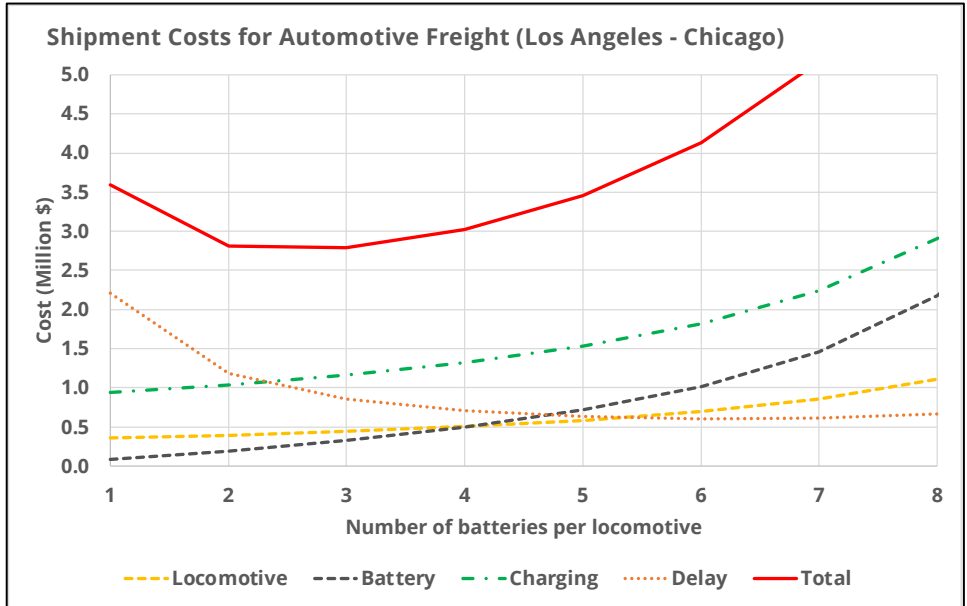

**Figure 4 – Costs of Shipments for Automotive Goods between Los Angeles and Chicago**

### 3.3 Coal Freight between Powder River Basin, Wyoming and Chicago

Relative to the previous two examples, coal freight involves much heavier yet slower shipments, resulting in distinct optimal results. Figure 5 shows the optimal number of batteries is one per locomotive, resulting in a 320-mile range and four stops over the 1,400-mile journey. The key trade-off shown by increasing the number of batteries is between the delay and fixed costs of batteries. However, as coal shipments are generally time-insensitive, the delay costs are minimal. Therefore, under these assumptions there would be no incentive to increase the number of battery tender cars to reduce the number of stops and associated delays, as seen in the monotonically increasing curve of the total cost. This illustrates an example of bulk commodities where time constraints play less of a role.

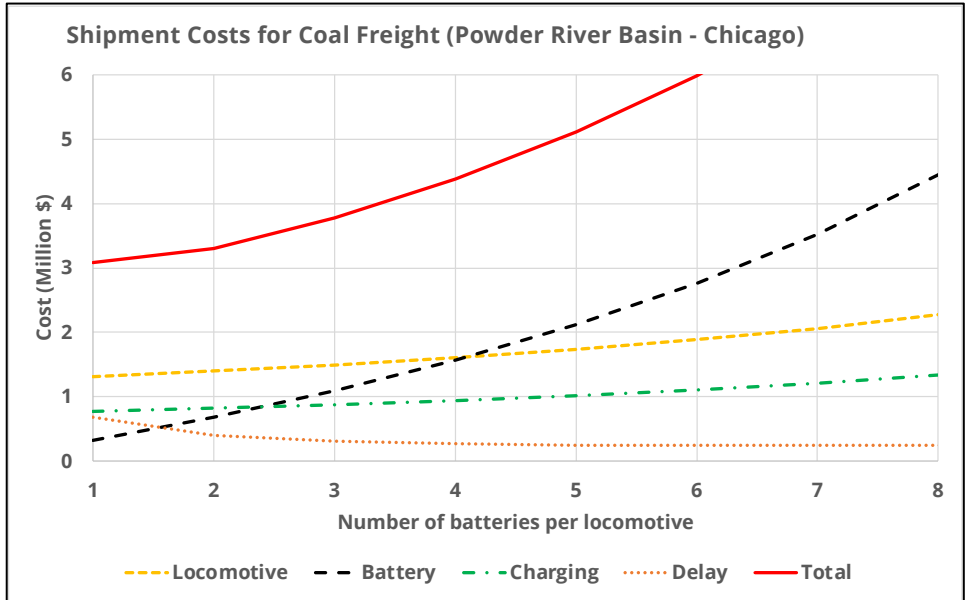

**Figure 5 - Costs of Shipments for Coal Freight between Powder River Basin and Chicago**

## 4    NETWORK-WIDE FREIGHT MARKET ANALYSIS

For a network-wide analysis of battery-electric tender configuration in the whole freight rail network, we apply the model to the American Class I railroads with rail freight shipment data from 2019 (see Appendix B for details). Flow data from the Carload Waybill Sample as reported by the railroads (Surface Transportation Board, 2019a) are aggregated into 22,501 markets of different origin-destination pairs, railroads, and commodity groups. The analysis is carried out with respect to eight commodity groups (intermodal, motor vehicles, forest products, agricultural & foods, chemical & petroleum, metal & ores, coal, and all others) and railroad regions (Western and Eastern). The differences in energy requirements and other shipment characteristics across commodities and regions are further discussed in Appendix B. The optimal number of batteries normalized per locomotive, $n^*/n_l$, is then calculated to minimize the total cost.

We first explore the scenarios when capital costs are included (following the formulation in Appendix A), then assess the results' sensitivity on unit delay costs and charging speeds, followed by the scenario excluding capital costs as formulated in Section 2. The results presented show median values consolidated by commodity groups, with error bars denoting standard deviation.

### 4.1    Inclusion of Capital Costs

Figure 6-Figure 8 show the optimal number of batteries, corresponding ranges, and number of stops per 1,000 miles for various commodity groups, with the columns showing the median values and error bars the standard deviation. The detailed results are also tabulated in Appendix C. The median of the optimal number of batteries lies between one and four, with optimal ranges of 150-350 miles. This suggests the adoption of multiple battery tender cars per locomotive can lead to lower costs for more than half of cases in intermodal, motor vehicles, forest products, and others.

There are considerable differences across commodity groups. Intermodal movements require the most batteries (four) due to high delay costs, as demonstrated in the linehaul example in Section 3.1. Other bulk commodities such as coal, chemicals, and agricultural products mostly require one battery per locomotive at optima, which suggests the delay reduction by additional batteries may not be worth their extra cost. Meanwhile, there is no substantial difference found across regions.

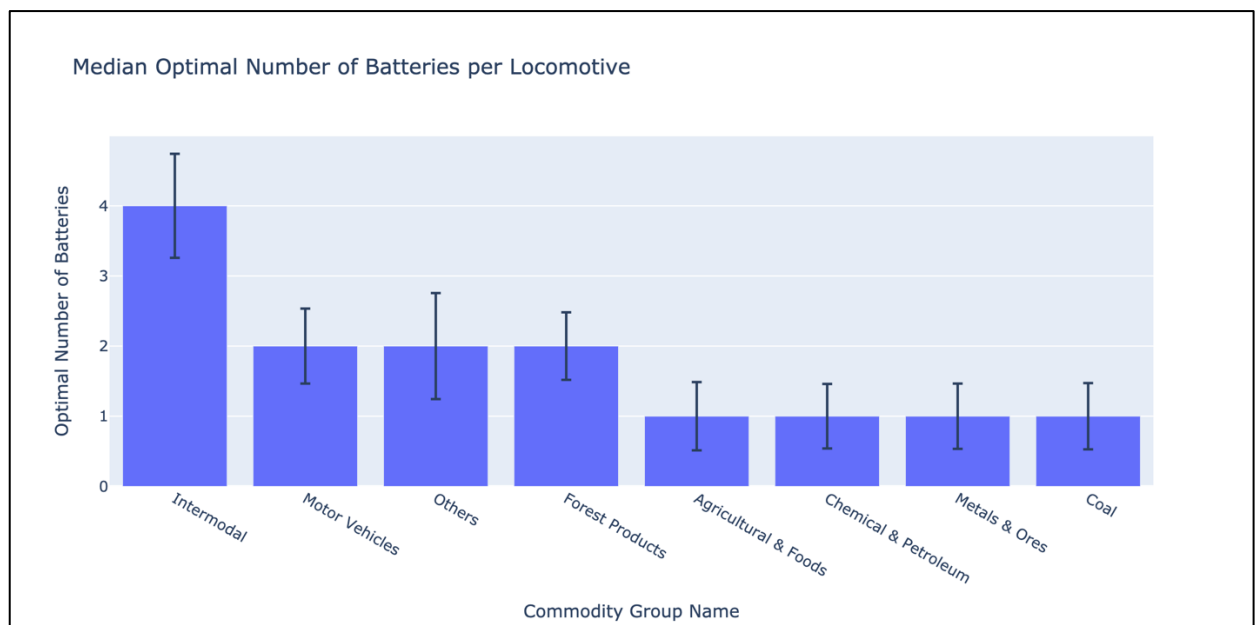

**Figure 6 – Median Optimal Number of Batteries per Locomotive for Various Commodity Groups**

When we compare across commodities, the difference in the number of batteries does not directly translate to differences in range. For example, both motor vehicles and forest products require two batteries per locomotive at the median as shown in Figure 6, but the range of the former is much lower than that of the latter, as shown in Figure 7. The underlying reason is the higher energy requirements per ton-mile of motor vehicles (see Table B.3 and further discussion in Appendix B), resulting in its lowest median range. Beyond this example, the corresponding locomotive ranges mostly fall between 200 and 300 miles despite the difference in the optimal number of batteries across commodities.

Nevertheless, coupling batteries increases the range for commodities like intermodal and motor vehicles. The possibility of a range increase may provide railroads with higher flexibility for charging facility deployment and train scheduling. Besides, it also reduces the gap in journey time with existing diesel, improving its competitiveness in some time-sensitive markets.

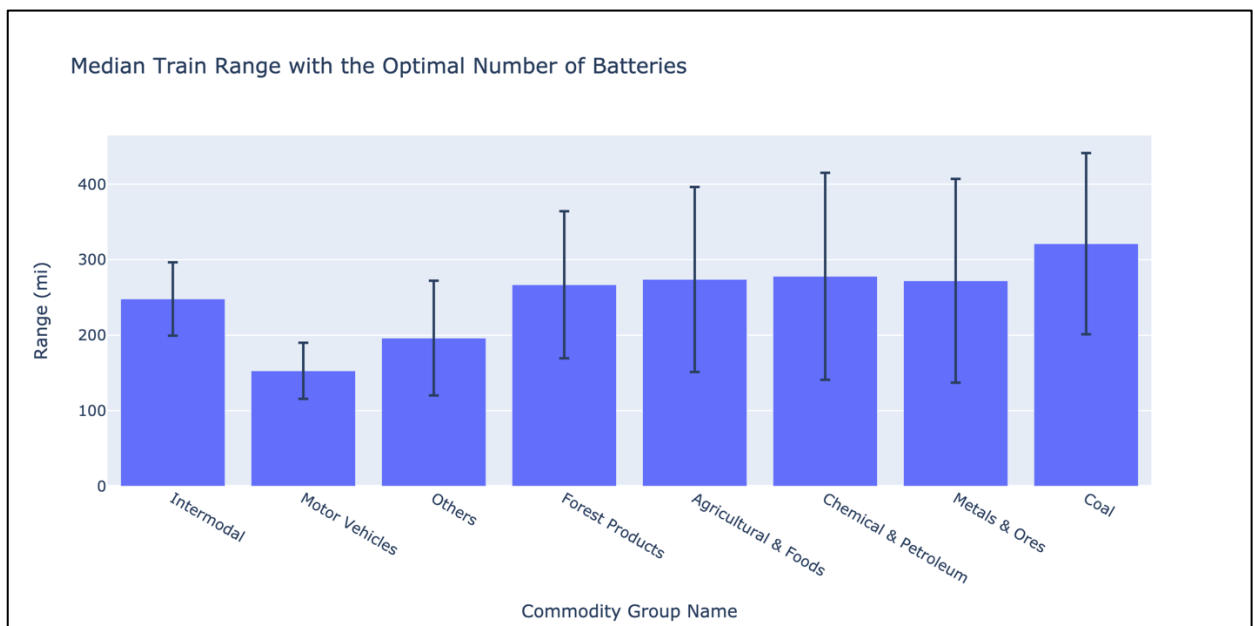

**Figure 7 – Median Train Range with the Optimal Number of Batteries**

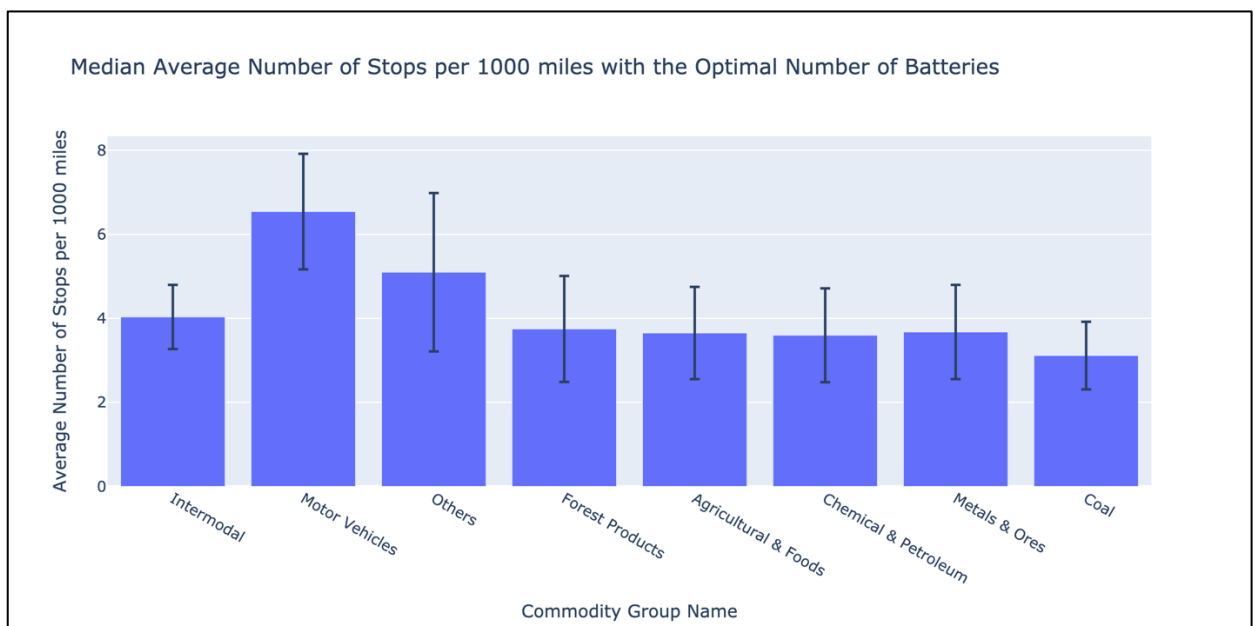

**Figure 8 – Median Average Number of Stops per 1,000 miles with the Optimal Number of Batteries**

Considering the scenario of deploying the optimal number of batteries per locomotive, Figure 9 further highlights the trade-off between the four cost components (holding – delay; fixed – locomotive, battery, electricity). For intermodal, delay costs account for more than 40% of total costs, underscoring the effects of the high time-value of goods. Electricity costs constitute considerable portions for both intermodal and motor vehicles due to their high energy requirements. The total costs of other commodities are primarily dominated by higher fixed (capital) costs in locomotives.

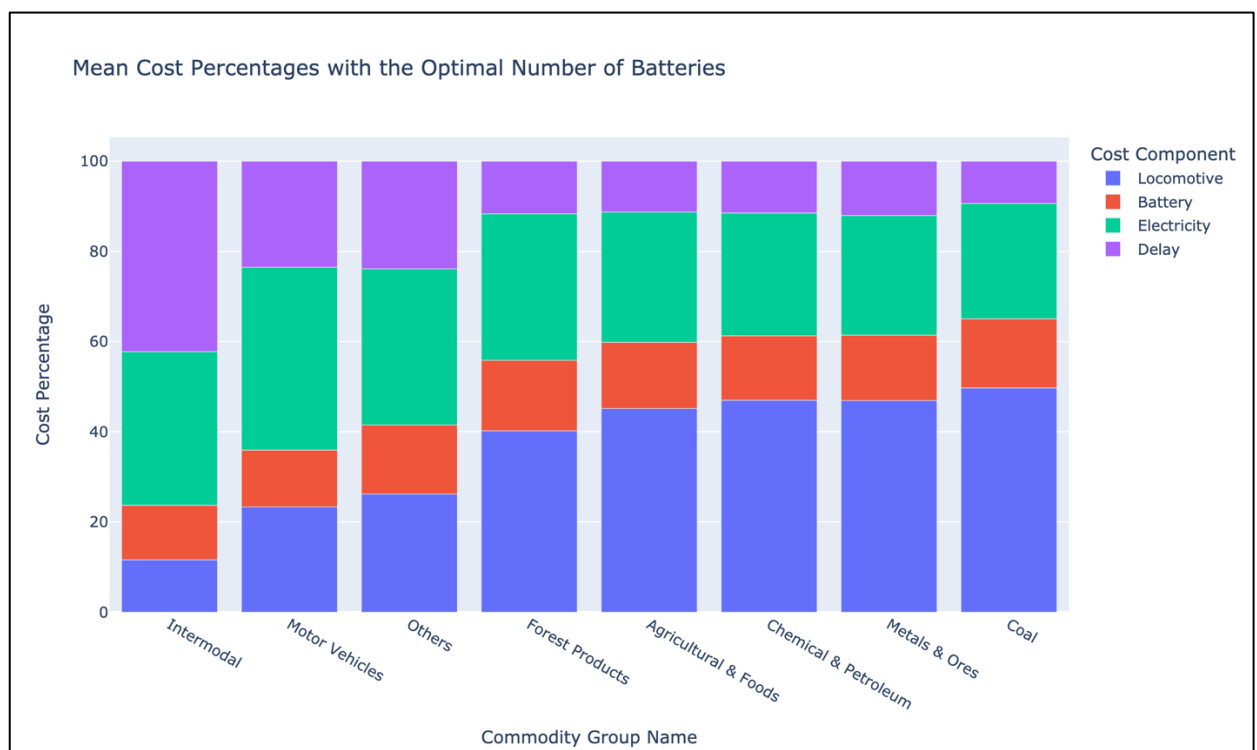

**Figure 9 – Mean Cost Component Percentages with the Optimal Number of Batteries**

Overall, more batteries deployed require more electricity to move, which may be balanced out by the better energy efficiency of battery-electric locomotives. This ultimately depends on the eventual cost of batteries and the cost of carbon emissions (or equivalently carbon tax). To minimize the economic costs of delay due to charging, financial costs may also be incurred to upgrade infrastructure or to speed up trains; otherwise, railroads may risk losing some time-sensitive freight orders.

## 4.2 Sensitivity Analysis: Unit Delay Costs

To evaluate the sensitivity of the previous results to the estimated delay costs, sensitivity analysis is carried out on various scales of unit delay costs (50%, 75%, 100%, 150%, and 200%). Results are shown in Figure 10-Figure 12.

As delay costs account for less than 30% of total costs in optimal cases (except for intermodal) from Figure 9, the optimal numbers of batteries do not show significant change to linear increases in unit delay costs. This can be explained by the considerable "fixed costs" incurred (capital costs of batteries) if an extra battery tender is added.

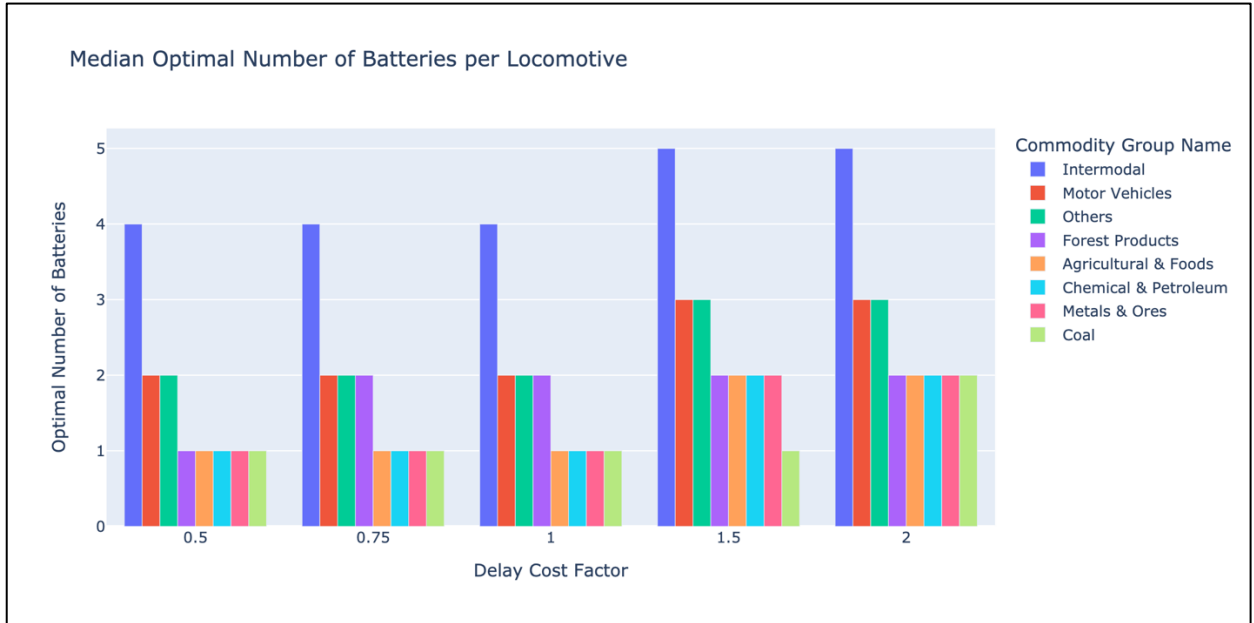

**Figure 10 – Median Optimal Number of Batteries per Locomotive for Different Delay Cost Factors**

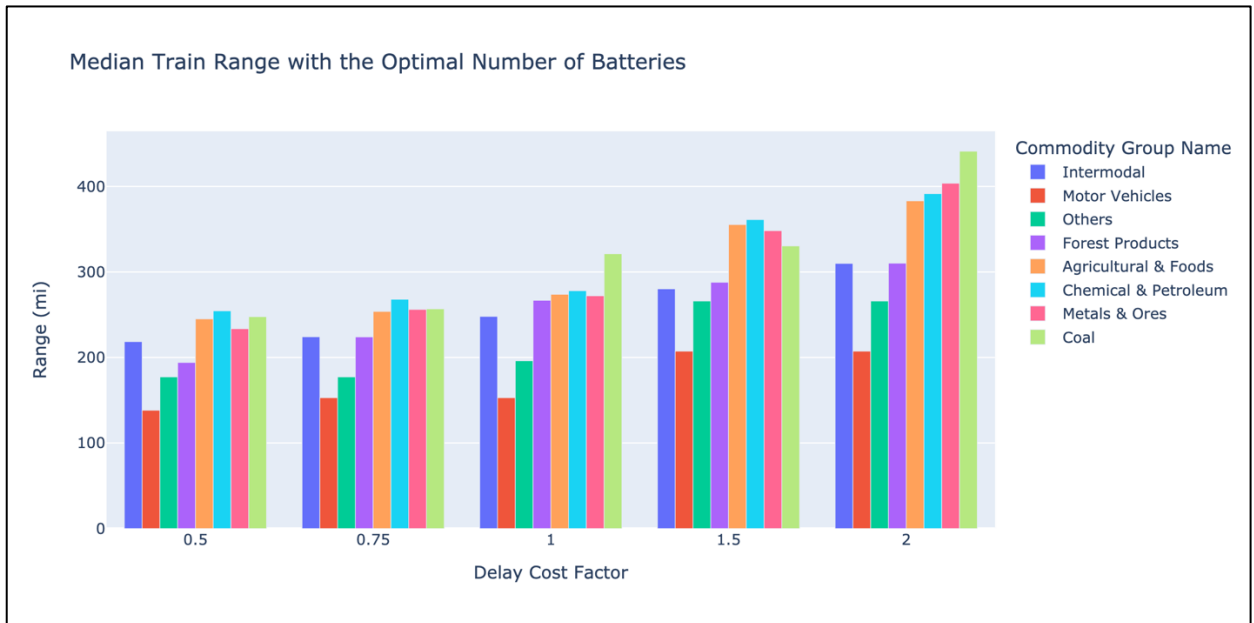

**Figure 11 – Median Train Ranges with the Optimal Number of Batteries under Different Delay Cost Factors**

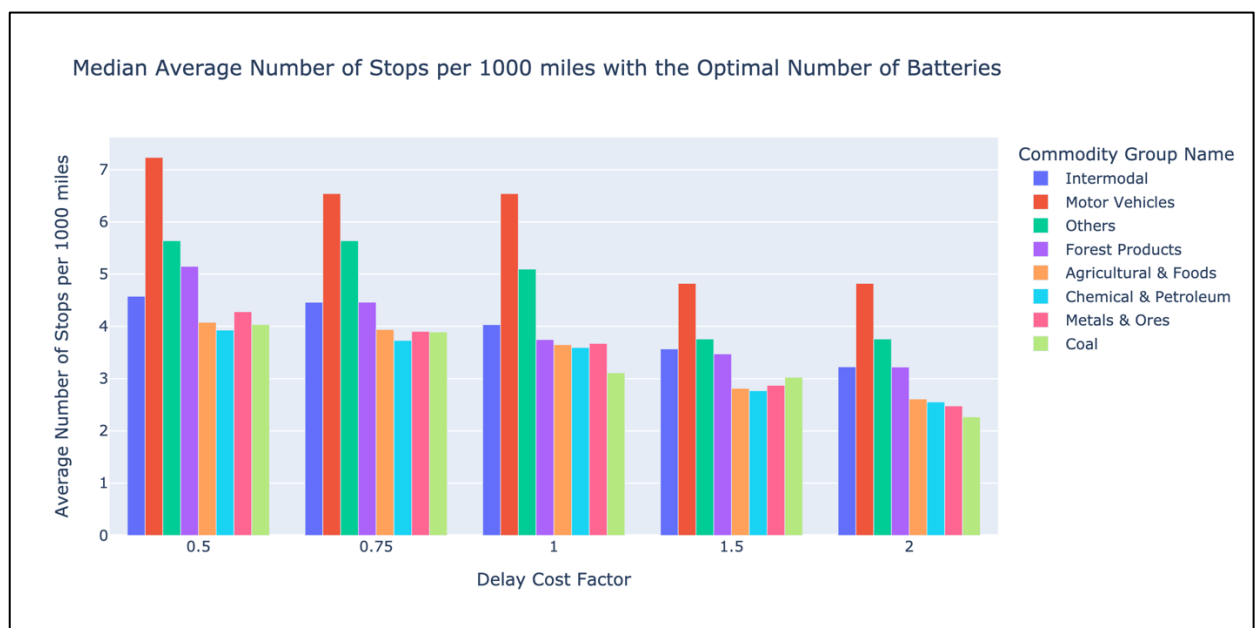

**Figure 12 - Median Average Number of Stops per 1,000 miles with the Optimal Number of Batteries under Different Delay Cost Factors**

## 4.3    Sensitivity Analysis: Charging Speed

Technological developments contribute to significant uncertainty in the cost parameters resulting from the techno-economic analysis. The proposed formulation enables sensitivity analysis of changes in such parameters, e.g., battery charging speeds. This analysis also considers the proposal of battery swapping for comparison under the assumption of constant charging cost and a stopping time of 0.5 hours.

Figure 13-Figure 15 show the optimal number of batteries, ranges, and number of stops per 1,000 miles for different charging speeds. Compared to the ambitious 3MW charger, currently available 400kW chargers would lead to many more batteries per locomotive at optimality, with seven for intermodal and three for coal, to compensate for the considerable charging time, as they are about seven times slower than their 3MW counterparts. Faster charging leads to lower ranges and more stops, since each charging stop brings less delay. With battery swapping technology, only one battery is needed per locomotive for most cases, except for intermodal.

In reality, the results should be subject to variations in charging costs, which are difficult to project based on current estimates. These cost forecasts are particularly uncertain for battery swapping, where the additional infrastructure costs may be offset by the cost savings offered by chargers with lower speeds.

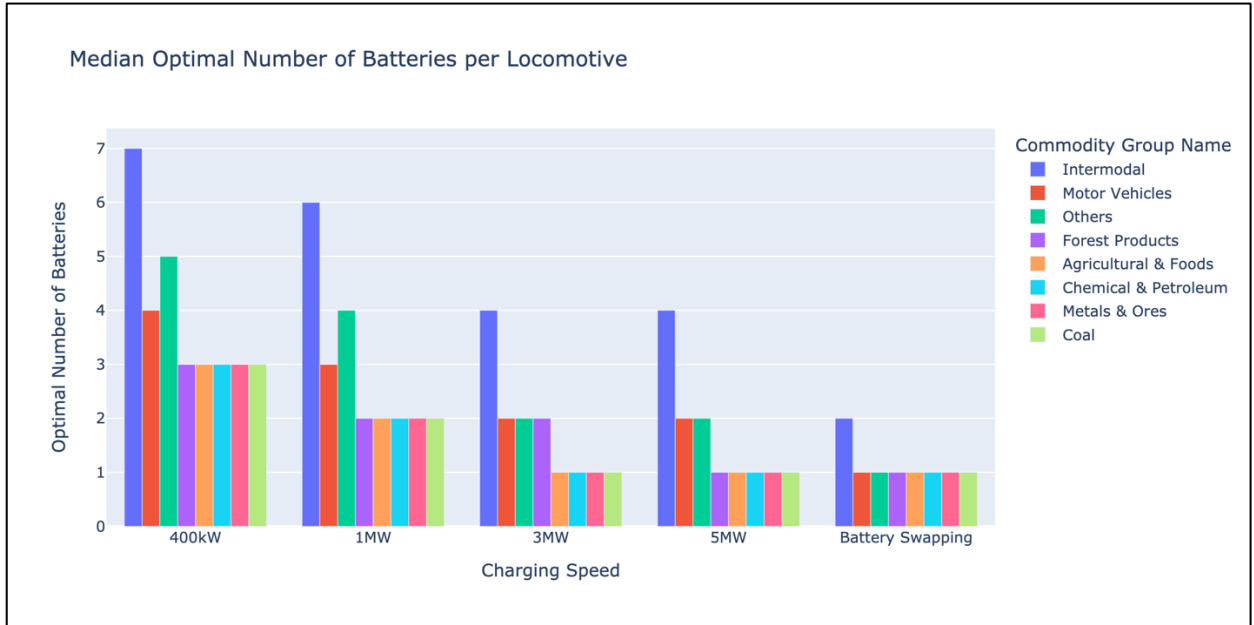

**Figure 13 – Median Optimal Number of Batteries per Locomotive under Different Charging Speeds**

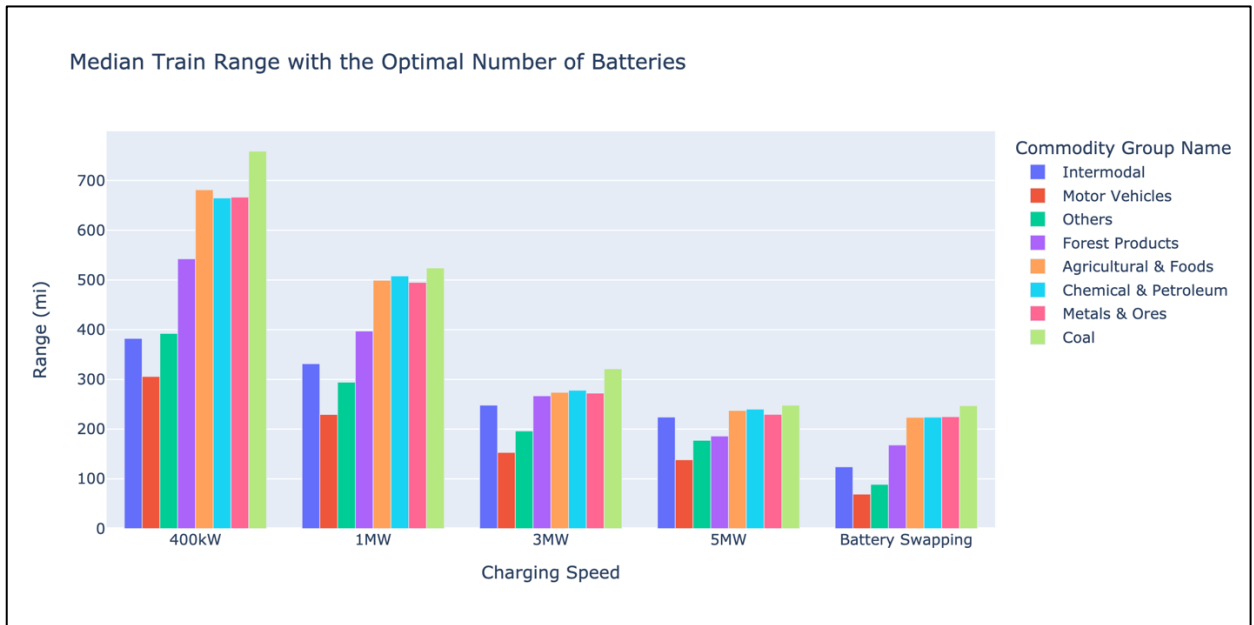

**Figure 14 – Median Train Range with the Optimal Number of Batteries under Different Charging Speeds**

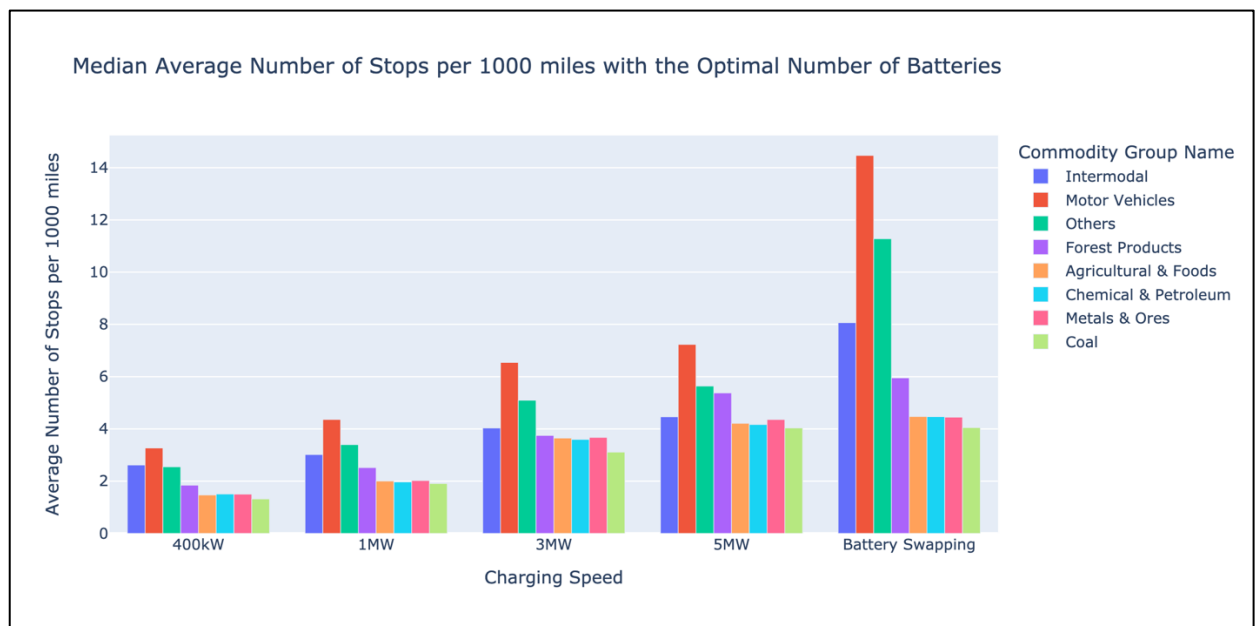

**Figure 15 – Median Average Number of Stops per 1,000 miles with the Optimal Number of Batteries under Different Charging Speeds**

## 4.4    Exclusion of Capital Costs

When capital costs for locomotives and batteries are irrelevant (see formulation in Section 2), the optimal number of batteries and resulting ranges increase. In Figure 16-Figure 18, the intermodal median case requires five batteries per locomotive (compared to four when capital costs are included), with other commodities around two to three (against one to two with capital costs). The resulting ranges are 300-800 miles, significantly higher than 150-350 miles in the earlier case with capital costs included. Similarly, the number of stops is generally reduced to one to five per 1,000 miles, in contrast to three to eight considering capital costs. The detailed results are also tabulated in Appendix C.

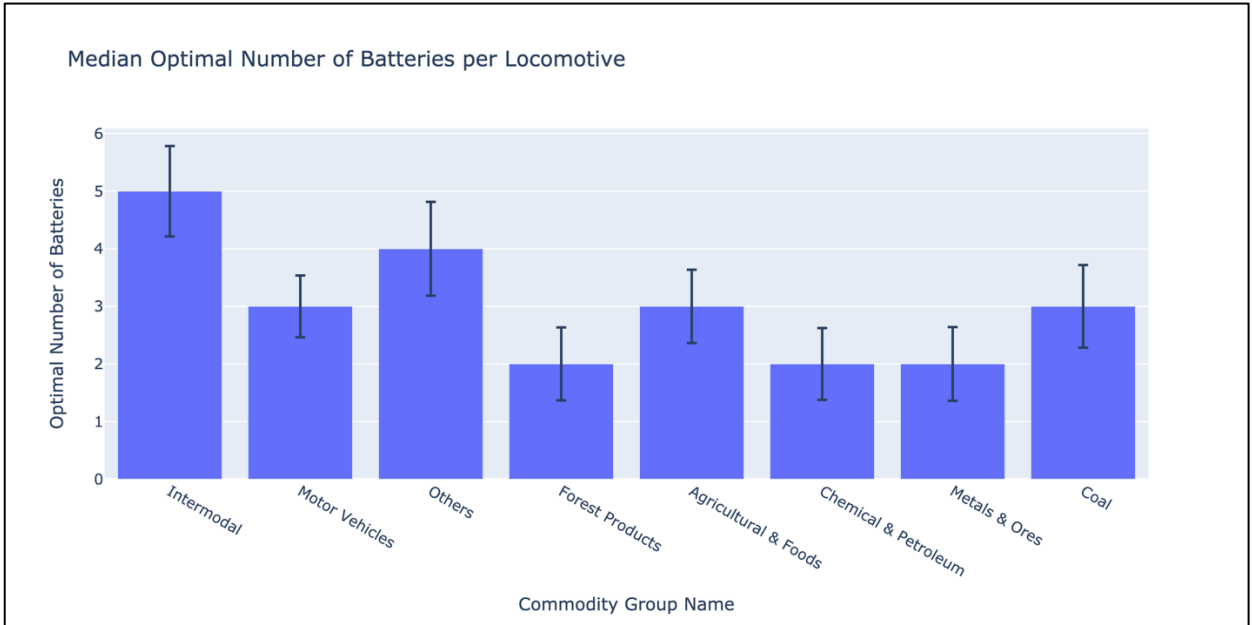

**Figure 16 – Median Optimal Number of Batteries per Locomotive for Various Commodity Groups under the Exclusion of Capital Costs**

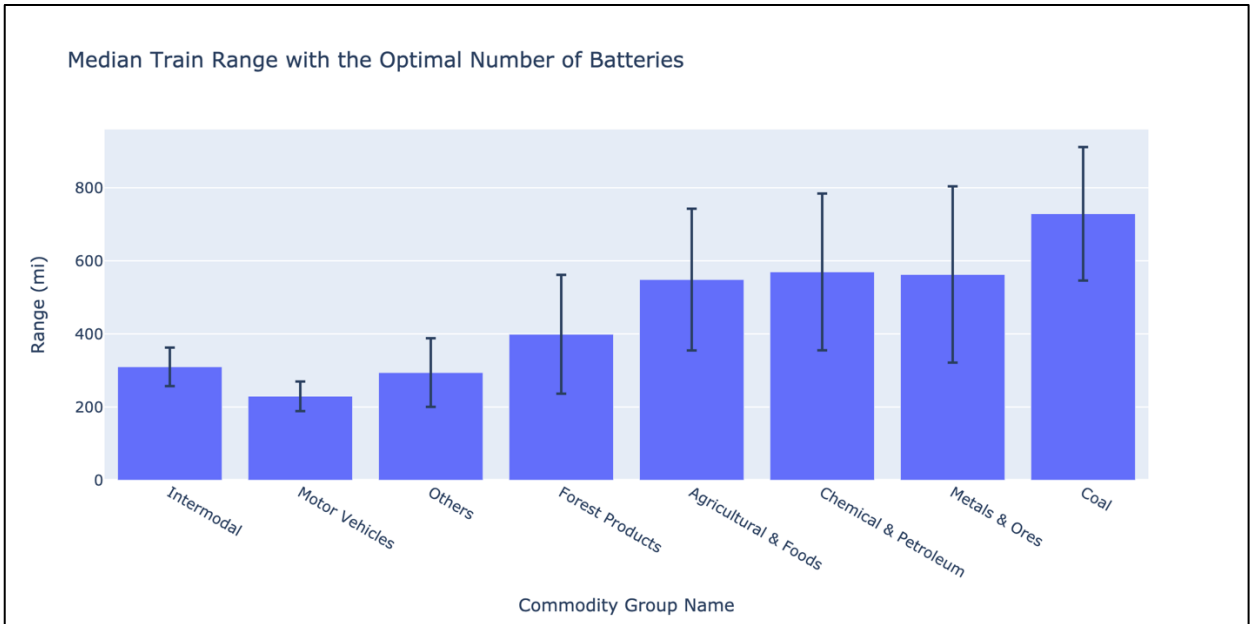

**Figure 17 – Median Train Ranges with the Optimal Number of Batteries under the Exclusion of Capital Costs**

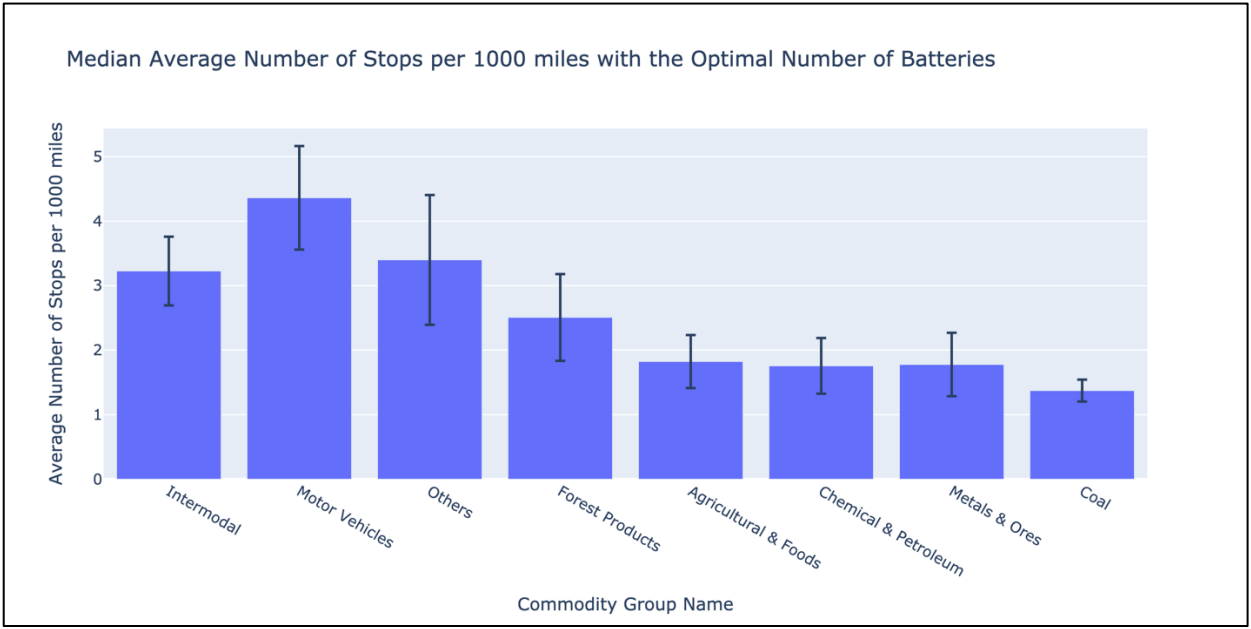

**Figure 18 - Median Average Number of Stops per 1,000 miles with the Optimal Number of Batteries under the Exclusion of Capital Costs**

Figure 19 illustrates the direct trade-off between delay and electricity costs, whereas the additional battery weight adds to the extra energy requirements. Electricity costs account for the major proportions of total costs at optima for most commodities except intermodal.

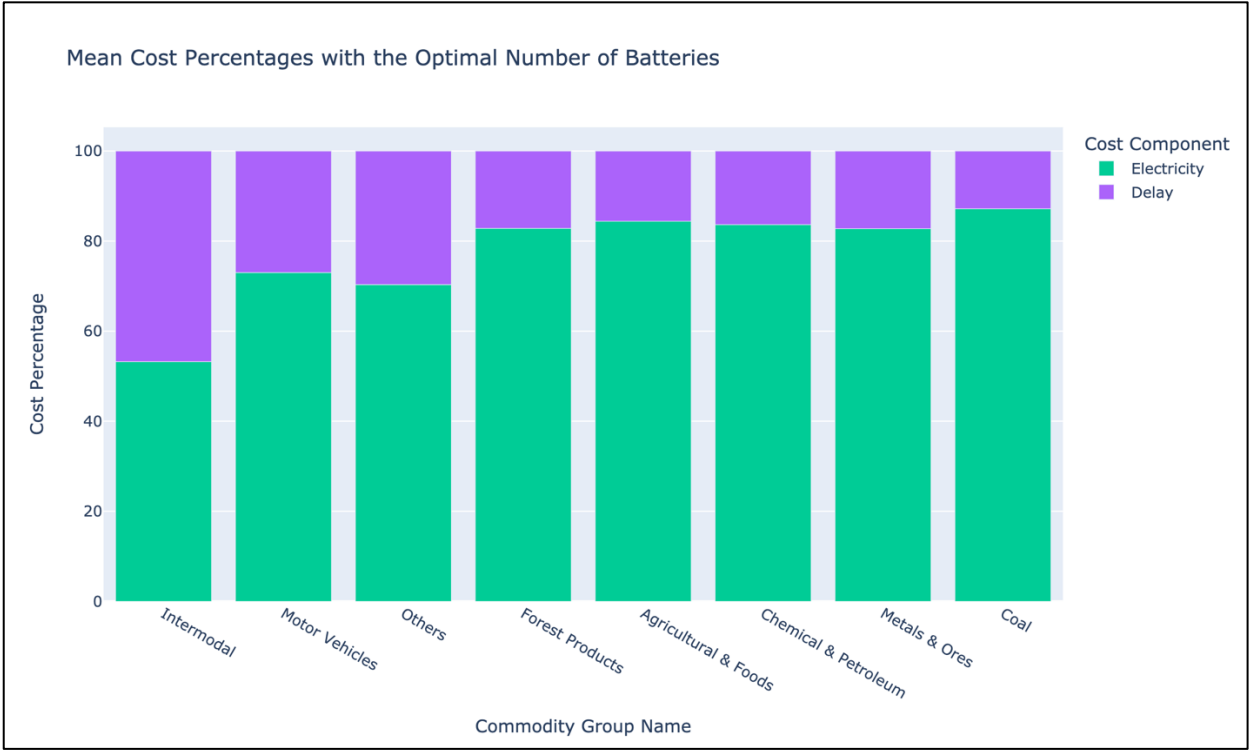

**Figure 19 - Mean Cost Component Percentages with the Optimal Number of Batteries under the Exclusion of Capital Costs**

## 5    CONCLUSION

We developed a convex optimization formulation to optimize energy storage tender car configuration in freight trains, i.e., to determine the number of energy storage tender cars per train, for different freight markets. With the consideration of holding costs for shipment delays and fixed costs for locomotive, battery, and charging equipment and operations, a concise formulation provides a tractable closed-form solution with a convex total cost function, which also bears a similar structure to an Economic Order Quantity (EOQ) model. These attractive solution properties are maintained under the full formulation in Appendix A.

The applicability of the model is then demonstrated with future battery-electric technology parameters for three numerical examples of linehaul rail freight and a network-wide freight market analysis. Results illustrate heterogeneity in optimal tender configurations with lighter yet more time-sensitive shipments (e.g., intermodal) favoring more battery tender cars. In contrast, for heavier commodities (e.g., coal) with lower values of time, single battery configurations are shown to be more economically sensible. The analysis also supports coupling multiple battery tender cars in many commodity markets across regions. Sensitivity analyses show that mild changes in delay costs do not greatly affect the optimal number of batteries; however, slower charging speeds lead to considerably more batteries at optima. Alternatively, battery swapping or other innovations that greatly reduce charging time would mean optimal single-battery configurations in many cases. The final scenario excluding capital costs leads to optima with many more batteries across all commodity groups.

By addressing a research gap for studying optimal tender arrangements for constrained on-board energy storage in freight rail to capture the operational impacts of delays, this framework enables industry analysts to assess the benefits of energy storage tenders, thereby preparing and planning for the roll-out of new energy technologies with limited shipment ranges. Along the line of previous studies initiated by U.S. ARPA-E, EU-RAIL, and other regions across the world looking into the deployment of alternative energy technologies in freight rail, this framework has natural decision-making applications for determining investments in locomotives, tender cars, and charging infrastructure, both in terms of quantity and time horizons. Moreover, the paper corroborates the benefits of coupling multiple battery tender cars in different rail freight markets from an economic perspective. Lastly, the framework formulates the best-case scenario for battery-electric tender car technologies with optimal battery tender car configurations. The associated emissions and costs can be compared with other energy technologies (e.g., hydrogen fuel cells) to assess their potential as economically sensible low-carbon alternatives to diesel and informing cost-benefit analyses and technology-related decisions in rail freight decarbonization investments.

Nevertheless, this paper is still subject to certain limitations which will benefit from further research:

1. The applicability of results is naturally limited by parameter uncertainties, in particular those regarding technological and cost forecasts. More sensitivity and robustness analyses would improve the optimal investment decision.
2. To achieve a tractable solution, this study relies on parameters to capture operational characteristics such as delay. A network-wide analysis can explicitly analyze these effects across routes, such as queuing and over-capacity.

3. This study focuses on energy technologies with critical storage constraints, such as battery-electric locomotives and does not consider cross-over technologies such as hybrid battery-diesel locomotives. The results can be used to compare with the cost and emissions reduction of conventional diesel operations and other alternatives including biofuel and e-fuel.

Adding to the freight rail decarbonization literature to analyze the economic, environmental, and operational impacts of alternative energy technologies, this study and further research in the area of energy storage tender cars can equip railroads and authorities with tools to evaluate options and select the best investment strategy. A similar model may also be applied to other transportation modes where the trade-off between energy storage and payload is critical, such as battery-electric planes. All these can strengthen the decarbonization efforts of the transportation sector to mitigate climate change.

## APPENDIX A  GENERALIZED MODEL WITH FIXED COSTS AS A FUNCTION OF ENERGY STORAGE TENDERS

We present a variation of the convex optimization model in Section 2 featuring fixed costs as a function of $n$ and show the updated total cost function remains convex and its individual cost components remain convex and monotonic. This generalization allows for a model specification which captures the time-dependent nature of certain fixed cost components, as opposed to assuming a distance-based or trip-based fixed cost.

Here, we represent the fixed costs, denoted as $k(n)$, varying in $n$ in (A.3)—in contrast with the constant fixed costs $k$ assumed in Section 2. In difference with Section 2, $n_l$ is the number of locomotives per train and $c_l$ and $c_n$ are, respectively, the equipment cost rates per unit of time for locomotives and for energy tender cars. Depending on a railroad's cost structure, $c_l$ and $c_n$ may also include costs associated with labor. The number of stops $s(n)$ in (A.1) and total trip duration $t(n)$ in (A.2) are the same as for the original model in Section 2. The fixed cost in (A.4) is composed of the per-train amortized equipment (and labor) costs and fuel/energy costs. As in Section 2, $f$ is the per-stop cost of refueling/recharging an energy tender car.

$$s(n) = \frac{D}{rn} \tag{A.1}$$

$$t(n) = t_0 + s(n) \cdot t_s \tag{A.2}$$

$$k(n) = (n_l \cdot c_l + n \cdot c_n)t(n) + f \cdot n \cdot s(n) \tag{A.3}$$

$$TC = k(n) \cdot \frac{Q}{L - \alpha n} + h \cdot t(n) \cdot Q \tag{A.4}$$

$$TC \equiv A\frac{n}{L - \alpha n} + B\frac{1}{Ln - \alpha n^2} + C\frac{1}{L - \alpha n} + E\frac{1}{n} + F \tag{A.5}$$

$$A = c_n t_0 Q \geq 0; \quad B = n_l c_l \frac{D}{r} t_s Q \geq 0; \quad C = \left(n_l c_l t_0 + c_n \frac{D}{r} t_s + f\frac{D}{r}\right)Q \geq 0;$$

$$E = h\frac{D}{r}t_s Q \geq 0; \quad F = ht_0 Q \geq 0 \tag{A.6}$$

To show the convexity of the total cost function (and thus guarantee the existence of an optimal solution), we re-group the total cost expression into five terms in (A.5)-(A.6) and make the assumptions in (A.7)-(A.9) on acceptable values for $n$ and $L$. These imply that $n \in [1, \frac{L-1}{\alpha}]$ and $L \in [2, \infty)$.

$n \geq 1 \Longrightarrow$ require at least one energy tender car per train $\tag{A.7}$
$L - \alpha n \geq 1 \Longrightarrow$ we must move at least one revenue carload per train $\tag{A.8}$
$L \geq 2 \Longrightarrow$ minimum total carloads of one energy tender car and one revenue carload $\tag{A.9}$

The first and second derivatives of the total cost function are shown in (A.10)-(A.11). All terms in the second derivative are positive, as $\left(\alpha n - \frac{L}{2}\right)^2 \geq 0$, $\frac{L^2}{4} \geq 0$, $n^3 \geq 0$, and $(L - \alpha n)^3 \geq 0$ based on the assumed domains of $n$ and $L$. Thus, as the entire second derivative is positive, the total cost function is shown to be convex. The optimal value of $n$ is shown in (A.12).

$$TC' = A\frac{L}{(L - \alpha n)^2} + B\frac{(2\alpha n - L)}{(Ln - \alpha n^2)^2} + C\frac{\alpha}{(L - \alpha n)^2} - E\frac{1}{n^2} \tag{A.10}$$

$$TC'' = A\frac{2\alpha L}{(L - \alpha n)^3} + B\frac{2\left(3\left(\alpha n - \frac{L}{2}\right)^2 + \frac{L^2}{4}\right)}{n^3(L - \alpha n)^3} + C\frac{2\alpha^2}{(L - \alpha n)^3} + E\frac{2}{n^3} \geq 0 \tag{A.11}$$

$$n^* = \frac{-\alpha(B + EL) + \sqrt{(B + EL)(AL^2 + B\alpha^2 + CL\alpha)}}{AL + c\alpha - E\alpha^2} \tag{A.12}$$

While the formulae above prove the convexity of the total cost, the following further evaluates the monotonicity and convexity of individual cost components.

In (A.13)-(A.18), take $C_L(n)$ to be the locomotive equipment cost component, $C_T(n)$ to be the energy tender car equipment cost component, $C_R(n)$ to be the refueling/charging cost component, $C_D(n)$ to be the delay cost component, and $C_K$ to be the constant cost component, we rearrange the terms of the total cost function into these groups and show each component is both monotonic and convex over the defined intervals for $n$ and $L$.

$$TC = C_L(n) + C_T(n) + C_R(n) + C_D(n) + C_K \tag{A.13}$$

$$C_L(n) = n_l c_l t_0 Q \frac{1}{L - \alpha n} \tag{A.14}$$

$$C_T(n) = c_n t_0 Q \frac{n}{L - \alpha n} \tag{A.15}$$

$$C_R(n) = f \frac{D}{r} Q \frac{1}{L - \alpha n} \tag{A.16}$$

$$C_D(n) = \left( n_l c_l \frac{1}{Ln - \alpha n^2} + c_n \frac{1}{L - \alpha n} + h \frac{1}{n} \right) \frac{D}{r} t_s Q \tag{A.17}$$

$$C_K = h t_0 Q \tag{A.18}$$

To show monotonicity, we show each of their first derivatives have constant sign for the defined $n$ in (A.19)-(A.22). Note that $(L - \alpha n)^2 \geq 0$ for the defined $n$ and $L$.

$$\frac{\partial C_L(n)}{\partial n} = n_l c_l t_0 Q \frac{\alpha}{(L - \alpha n)^2} > 0 \implies C_L(n) \text{ monotonically increasing} \tag{A.19}$$

$$\frac{\partial C_T(n)}{\partial n} = c_n t_0 Q \frac{L}{(L - \alpha n)^2} > 0 \implies C_T(n) \text{ monotonically increasing} \tag{A.20}$$

$$\frac{\partial C_R(n)}{\partial n} = f \frac{D}{r} Q \frac{\alpha}{(L - \alpha n)^2} > 0 \implies C_R(n) \text{ monotonically increasing} \tag{A.21}$$

$$\frac{\partial C_D(n)}{\partial n} = \left( n_l c_l \frac{(2\alpha n - L)}{(Ln - \alpha n^2)^2} + c_n \frac{\alpha}{(L - \alpha n)^2} - h \frac{1}{n^2} \right) \frac{D}{r} t_s Q \leq 0$$

$$\implies C_D(n) \text{ monotonically decreasing for} \tag{A.22}$$

$$c_n \leq n_l c_l \frac{(L - 2\alpha n)}{\alpha n^2} + h(L - \alpha n)^2; \quad L - 2\alpha n \leq 0$$

To show convexity, we show each of their second derivatives are non-negative in (A.23)-(A.26).

$$\frac{\partial^2 C_L(n)}{\partial n^2} = n_l c_l t_0 Q \frac{2\alpha^2}{(L - \alpha n)^3} \geq 0 \tag{A.23}$$

$$\frac{\partial^2 C_T(n)}{\partial n^2} = c_n t_0 Q \frac{2\alpha L}{(L - \alpha n)^3} \geq 0 \tag{A.24}$$

$$\frac{\partial^2 C_R(n)}{\partial n^2} = f \frac{D}{r} Q \frac{2\alpha^2}{(L-\alpha n)^3} \geq 0 \tag{A.25}$$

$$\frac{\partial^2 C_D(n)}{\partial n^2} = 2\left(n_l c_l \frac{(3\alpha^2 n^2 - 3L\alpha n + L^2)}{(Ln - \alpha n^2)^3} + c_n \frac{\alpha^2}{(L-\alpha n)^3} + h\frac{1}{n^3}\right)\frac{D}{r} t_s Q \geq 0 \tag{A.26}$$

Similar to the main results presented in Section 2, due to the convexity of the total cost function, to obtain an integer number of energy storage tender cars, we only need to check the two adjacent integer values.

## APPENDIX B  ANALYSIS DATA SOURCES AND PARAMETERS

Table B.1 summarizes the parameter values utilized in the applications to follow.

Technical parameters for battery-electric technology (#1-#12) and the original payload (#27) are referenced from (Popovich et al., 2021). Electricity costs (#13), emissions (#14), and emission costs (#15) are referenced from (M. Wang et al., 2021), (U.S. Energy Information Administration, 2021), and (Carleton and Greenstone, 2021) respectively. Locomotive data (#16-#20) are calculated based on Annual Report of Finances and Operations ("R-1 Report") submitted by Class I railroads to the Surface Transportation Board (STB) (Surface Transportation Board, 2019b). The cost of capital rate for railroads (#21) is advised in (Surface Transportation Board, 2020). Other train operations data (#23, #24, and #25) are also referenced from railroad submissions (Surface Transportation Board, 2019a), (Surface Transportation Board, 2019c), and (Surface Transportation Board, 2019b), respectively. Diesel cost (#28) and energy characteristics (#29 and #30) are referenced respectively from (Kelly et al., 2022) and (M. Wang et al., 2021).

**Table B.1 – Data Source and Parameter Values**

| # | Parameter | Value |
|---|---|---|
| 1. | Unit weight of battery tender car (ton) | 150 |
| 2. | Battery capacity (MWh) | 14 |
| 3. | Charging speed (MW) | 3 |
| 4. | Charging depth | 80% |
| 5. | Battery energy efficiency | 95% |
| 6. | Capital cost of battery + inverter + boxcar ($) | 1,271,816 |
| 7. | Future cost of battery ($) | 452,908 |
| 8. | Battery maintenance cost ($/day) | 100 |
| 9. | Battery lifetime (years) | 13 |
| 10. | Relative energy efficiency of battery-electric to diesel | 2.44 |
| 11. | Discount rate | 3% |
| 12. | Time horizon (years) | 26 |
| 13. | Charging cost ($/kWh) | 0.15 |

| # | Parameter | Value |
|---|-----------|-------|
| 14. | Electricity grid carbon emission (kg $CO_2$ eqv/kWh) | 0.387 |
| 15. | Unit cost of carbon emissions ($/ton $CO_2$ eqv) | 125 |
| 16. | Locomotive utilization by road service hour | 25% |
| 17. | Five-year average new locomotive cost ($) | 2,560,000 |
| 18. | Five-year average annual capital cost ($) | 389,000 |
| 19. | Five-year average annual operating expense (admin, repair, maintenance, etc.) ($) | 127,000 |
| 20. | Five-year average annual total cost ($) | 516,000 |
| 21. | Cost of capital rate | 9.34% |
| 22. | Time horizon (years) | 20 |
| 23. | Freight demand | (Various by origin-destination by railroad by commodity) |
| 24. | Train speed (mph) | (See Table B.2 - various by railroad by train type) |
| 25. | Energy requirement (BTU/ton-mile) | (See Table B.3 - various by railroad by commodity) |
| 26. | Original stopping time (h) | 4 |
| 27. | Nominal payload (ton) | 1,700 |
| 28. | Diesel cost ($/gallon) | 2.47 |
| 29. | Diesel lower heating value (BTU/gallon) | 129,488 |
| 30. | Diesel emissions (kg CO2 eqv/gallon) | 12.36 |

**Table B.2 - Average Train Speed for Various Train Types and Railroads (in mph)**

| Railroad | Western | | | | Eastern | | |
|----------|---------|------|------|------|---------|------|------|
| **Train Type** | **BNSF** | **CN** | **CP** | **UP** | **CSX** | **KCS** | **NS** |
| **Intermodal** | 32.1 | 24.9 | 32.6 | 29.1 | 31.5 | 31.9 | 27.5 |
| **Grain unit** | 22.2 | 23.5 | 24.5 | 23.3 | 22.2 | 24.6 | 19.1 |
| **Coal unit** | 19.8 | 23.3 | 23.1 | 25.2 | 21.9 | 24.6 | 18.6 |
| **Automotive unit** | 25.4 | N/A | 26.5 | 23.6 | N/A | N/A | 21.2 |
| **Crude oil unit** | 21.9 | 23 | 25.4 | 20.8 | 28.6 | N/A | 19.4 |
| **Ethanol unit** | 21 | 25.5 | 24.4 | 20.8 | 26.9 | N/A | 20.3 |
| **Other Unit** | N/A | 21.2 | N/A | N/A | N/A | N/A | N/A |
| **Manifest** | 20.3 | 21.4 | 23.7 | 21 | 23.8 | 25.8 | 20.3 |
| **System** | 23.7 | 22.3 | 25.1 | 23.8 | 25.3 | 26.4 | 22.3 |

The computation of specific parameters is further discussed as follows.

## Flow Data

Data on the U.S. Class I railroads, defined as those with annual revenues of over $500,000,000 by the U.S. Surface Transportation Board, is aggregated at the regional level into "Western" (BNSF

Railway (BNSF), Canadian National Railway (CN), Canadian Pacific Railway (CP), and Union Pacific Railroad (UP)) and "Eastern" (CSX Transportation (CSX), Kansas City Southern Railway (KCS), and Norfolk Southern Railway (NS)). The original payload per locomotive is replaced with a nominal payload (1,700 tons in consistency with (Popovich et al., 2021)) if the recorded payload is too low.

**Energy Requirement, Range, Charging, and Emission Cost**

Battery-electric locomotive energy requirements are estimated based on the current diesel fuel consumption by Class I railroads which is then adjusted by the relative energy efficiency of battery-electric to diesel powertrains (Popovich et al., 2021). The average diesel fuel consumption (in gal/ton-mile) for individual railroads in 2019 is evaluated from (M. Wang et al., 2021). Diesel fuel consumption depends on a number of factors in trip characteristics (e.g., speed, terrain, and congestion) and train equipment (e.g., locomotive, car types, operating modes) (Fullerton et al., 2015; Heinold, 2020; ICF International, 2009; Liu et al., 2015; Tolliver and Lu, 2020). Factors from (M. Wang et al., 2021) are applied to improve the accuracy of estimating the energy requirements for the nine primary commodity groups as recorded by the American Association of Railroads (AAR) (2021b), which are summarized in Table B.3.

**Table B.3 - Estimated Diesel Energy Requirement for Various Commodities and Railroads**

| Commodity | Diesel Energy Requirement (BTU/ton-mile) | |
|---|---|---|
| | Western | Eastern |
| Agricultural & Foods | 155 | 155 |
| Chemical & Petroleum | 153 | 153 |
| Coal | 109 | 109 |
| Forest Products | 223 | 224 |
| Intermodal | 893 | 893 |
| Metals and Ores | 155 | 155 |
| Motor Vehicles | 724 | 725 |
| Nonmetallic Products | 130 | 131 |
| Others | 564 | 565 |

The locomotive range (in ton-miles) is evaluated as the effective battery capacity (in MWh) divided by the energy requirement (kWh/ton-mile) of specific commodities and railroads. The effective battery capacity is the assumed capacity adjusted by the charging depth and battery energy efficiency.

The charging cost (in $/kWh) estimated in (M. Wang et al., 2021) is taken as $0.15/kWh. The carbon emissions (in tons of $CO_2$ equivalent) can be estimated from the energy consumed at charging stations (in MWh) referencing the carbon intensity of the U.S. electricity grid (U.S. Energy Information Administration, 2021). It should be noted that both the cost of charging and the carbon intensities of power generation vary greatly across states due to differences in the generation mix. Afterward, the economic cost of carbon emissions is evaluated based on (Carleton and Greenstone, 2021) and added to the total charging cost.

**Marginal Costs of Batteries and Locomotives**

The hourly marginal cost of batteries and locomotives are estimated as the respective equivalent uniform cost rates amortized over their service lives, following a similar methodology as in (Popovich et al., 2021). The current and future capital costs and recurring maintenance costs (Surface Transportation Board, 2019b) are discounted as continuous cash flows, which are then used to estimate the hourly marginal cost with reference to the locomotive utilization by road services hour. The estimated marginal costs per road services hour are $58 for a battery and $236 for a locomotive.

**Delay Costs**

The delay costs estimated by Lovett et al. (2015) captured the costs imposed on railroads (crew, locomotives, fuel, railcars, and lading), shippers (inventory devaluation and holding), and the public (emissions and level crossings). The values based on the price level in 2015 were converted to 2019 values with the producer price index in the rail sector (U.S. Bureau of Labor Statistics, 1996) and are summarized in Table B.4 per train car for different trip distances for unit, manifest, and intermodal trains.

**Table B.4 – Hourly Delay Cost per Train Car (in 2019 USD)**

| Trip distance (mi) | 0-1,000 | 1,000-1,500 | >1,500 |
|---|---|---|---|
| **Unit train** | 8.42 | 8.42 | 8.42 |
| **Manifest train** | 17.57 | 17.57 | 17.57 |
| **Intermodal train** | 26.06 | 26.95 | 28.36 |

**APPENDIX C  DETAILED RESULT TABLES**

**Numerical Examples – Linehaul Shipments**

The detailed results of the numerical examples in Section 4 are shown in Table C.1.

To demonstrate how the costs can be compared to the conventional diesel locomotives, the table also shows their costs consisting of locomotive and fuel (including emissions costs) for each example. Such comparison relies on the current battery technology cost forecast, delay cost estimate, and emissions cost evaluation. The battery-electric technology offers lower financial costs (sum of locomotives, batteries, and charging) due to the improvement in energy efficiency, aligning with previous literature (Popovich et al., 2021). However, it becomes less attractive with higher total costs when the considerable delay costs are considered.

The model can be used to outline the frontier of technological advancement required for battery technology to be beneficial after considering the optimal battery arrangement. It also highlights the potential benefits brought by technologies (e.g., battery swapping) that reduces the charging time required.

**Table C.1 – Detailed Results of the Numerical Examples in Section 3**

| Batteries per Locomotive, $n/n_l$ | Locomotives, $n_l$ | Range (mile), $R(n)$ | Stops, $s(n)$ | Journey time, $t(n)$ (h) | Cost ($) | | | | |
|---|---|---|---|---|---|---|---|---|---|
| | | | | | Locomotive, $C_L(n)$ | Battery, $C_T(n)$ | Charging, $C_R(n)$ | Delay, $C_D(n)$ | Total, $TC$ |
| 1 | 28 | 62 | 37 | 224 | 496,989 | 121,578 | 2,313,149 | 15,498,078 | 18,429,794 |
| 2 | 31 | 124 | 19 | 150 | 550,238 | 269,209 | 2,560,987 | 7,946,048 | 11,326,481 |
| 3 | 35 | 186 | 12 | 125 | 616,266 | 452,271 | 2,868,305 | 5,460,226 | 9,397,068 |
| 4 | 39 | 248 | 9 | 113 | 700,303 | 685,259 | 3,259,438 | 4,250,628 | 8,895,627 |
| 5 | 45 | 310 | 7 | 105 | 810,877 | 991,822 | 3,774,086 | 3,564,142 | 9,140,926 |
| 6 | 54 | 372 | 6 | 100 | 962,916 | 1,413,346 | 4,481,727 | 3,157,623 | 10,015,612 |
| 7 | 66 | 434 | 5 | 97 | 1,185,127 | 2,029,420 | 5,515,971 | 2,941,430 | 11,671,948 |
| 8 | 86 | 496 | 5 | 94 | 1,540,666 | 3,015,138 | 7,170,763 | 2,902,605 | 14,629,171 |
| 9 | 123 | 558 | 4 | 92 | 2,200,951 | 4,845,758 | 10,243,947 | 3,122,963 | 20,413,618 |
| 10 | 216 | 620 | 4 | 90 | 3,851,664 | 9,422,307 | 17,926,907 | 4,032,123 | 35,233,001 |

| | Locomotives, $n_l$ | | | Journey time, $t(n)$ (h) | Cost ($) | | | | |
|---|---|---|---|---|---|---|---|---|---|
| | | | | | Locomotive, $C_L(n)$ | | Fuel, $C_R(n)$ | | Total, $TC$ |
| Diesel | 25 | | | 76 | 453,137 | | 2,571,040 | | 3,024,177 |

| **Batteries per Locomotive, $n/n_l$** | **Locomotives, $n_l$** | **Range (mile), $R(n)$** | **Stops, $s(n)$** | **Journey time, $t(n)$ (h)** | **Cost ($)** | | | | |
|---|---|---|---|---|---|---|---|---|---|
| | | | | | **Locomotive, $C_L(n)$** | **Battery, $C_T(n)$** | **Charging, $C_R(n)$** | **Delay, $C_D(n)$** | **Total, $TC$** |
| 1 | 14 | 76 | 30 | 138 | 358,112 | 87,605 | 938,090 | 2,212,018 | 3,595,825 |
| 2 | 15 | 153 | 15 | 139 | 396,481 | 193,982 | 1,038,600 | 1,185,905 | 2,814,969 |
| 3 | 17 | 229 | 10 | 141 | 444,059 | 325,890 | 1,163,232 | 856,651 | 2,789,832 |
| 4 | 20 | 306 | 8 | 144 | 504,612 | 493,772 | 1,321,855 | 705,534 | 3,025,773 |
| 5 | 23 | 382 | 6 | 147 | 584,288 | 714,671 | 1,530,569 | 630,791 | 3,460,318 |
| 6 | 27 | 459 | 5 | 151 | 693,842 | 1,018,406 | 1,817,550 | 601,701 | 4,131,499 |
| 7 | 33 | 535 | 4 | 157 | 853,960 | 1,462,326 | 2,236,985 | 611,005 | 5,164,275 |
| 8 | 43 | 612 | 4 | 167 | 1,110,147 | 2,172,599 | 2,908,080 | 667,995 | 6,858,821 |
| 9 | 62 | 688 | 3 | 186 | 1,585,925 | 3,491,676 | 4,154,400 | 813,932 | 10,045,933 |
| 10 | 108 | 764 | 3 | 232 | 2,775,369 | 6,789,370 | 7,270,200 | 1,227,897 | 18,062,836 |
| | **Locomotives, $n_l$** | | | **Journey time, $t(n)$ (h)** | **Cost ($)** | | | | |
| | | | | | **Locomotive, $C_L(n)$** | | **Fuel, $C_R(n)$** | | **Total, $TC$** |
| Diesel | 13 | | | 109 | 326,514 | | 1,042,610 | | 1,369,123 |

**(b) Section 4.2 Automotive Goods between Los Angeles and Chicago**

| | | | | | Cost ($) | | | | |
|---|---|---|---|---|---|---|---|---|---|
| **Batteries per Locomotive, $n/n_l$** | **Locomotives, $n_l$** | **Range (mile), $R(n)$** | **Stops, $s(n)$** | **Journey time, $t(n)$ (h)** | **Locomotive, $C_L(n)$** | **Battery, $C_T(n)$** | **Charging, $C_R(n)$** | **Delay, $C_D(n)$** | **Total, $TC$** |
| 1 | 79 | 320 | 4 | 88 | 1,314,845 | 321,650 | 772,193 | 677,627 | 3,086,315 |
| 2 | 84 | 640 | 2 | 79 | 1,399,130 | 684,537 | 821,692 | 394,151 | 3,299,510 |
| 3 | 90 | 960 | 1 | 77 | 1,494,961 | 1,097,134 | 877,973 | 304,713 | 3,774,780 |
| 4 | 96 | 1280 | 1 | 75 | 1,604,884 | 1,570,408 | 942,529 | 264,620 | 4,382,442 |
| 5 | 104 | 1600 | 1 | 74 | 1,732,256 | 2,118,805 | 1,017,333 | 245,146 | 5,113,540 |
| 6 | 113 | 1920 | 1 | 74 | 1,881,588 | 2,761,752 | 1,105,034 | 236,970 | 5,985,345 |
| 7 | 123 | 2240 | 1 | 73 | 2,059,097 | 3,526,011 | 1,209,283 | 236,416 | 7,030,806 |
| 8 | 136 | 2560 | 1 | 73 | 2,273,586 | 4,449,489 | 1,335,250 | 242,070 | 8,300,395 |
| 9 | 152 | 2880 | 0 | 73 | 2,537,956 | 5,587,731 | 1,490,512 | 253,745 | 9,869,944 |
| 10 | 172 | 3200 | 0 | 72 | 2,871,898 | 7,025,510 | 1,686,632 | 272,220 | 11,856,260 |

**(c) Section 4.3 Coal Freight between Powder River Basin, Wyoming and Chicago**

| | **Locomotives, $n_l$** | | | **Journey time, $t(n)$ (h)** | Cost ($) | | | | |
|---|---|---|---|---|---|---|---|---|---|
| | | | | | **Locomotive, $C_L(n)$** | | **Fuel, $C_R(n)$** | | **Total, $TC$** |
| Diesel | 74 | | | 71 | 1,240,138 | | 871,281 | | 2,111,418 |

5 **Network-wide Freight Market Analysis**
6
7 The detailed results of the numerical examples in Section 5 are shown in Table C.2.
8

**Table C.2 - Detailed Results of the Network-wide Freight Market Analysis in Section 4**

| (a) Section 5.1 Inclusion of Capital Costs | | | | | | | | |
|---|---|---|---|---|---|---|---|---|
| Commodity Group Name | | Agricultural & Foods | Chemical & Petroleum | Coal | Forest Products | Intermodal | Metals & Ores | Motor Vehicles | Others |
| Number of Batteries per Locomotives | Mean | 1.38 | 1.29 | 1.31 | 1.65 | 4.35 | 1.31 | 2.27 | 2.54 |
| | Standard Deviation | 0.49 | 0.46 | 0.47 | 0.48 | 0.74 | 0.47 | 0.53 | 0.76 |
| | 25th Percentile | 1 | 1 | 1 | 1 | 4 | 1 | 2 | 2 |
| | Median | 1 | 1 | 1 | 2 | 4 | 1 | 2 | 2 |
| | 75th Percentile | 2 | 2 | 2 | 2 | 5 | 2 | 2 | 3 |
| Range (mi) | Mean | 313 | 313 | 339 | 279 | 256 | 316 | 163 | 216 |
| | Standard Deviation | 123 | 137 | 120 | 97 | 49 | 135 | 37 | 76 |
| | 25th Percentile | 229 | 216 | 251 | 200 | 224 | 221 | 138 | 173 |
| | Median | 274 | 278 | 321 | 267 | 248 | 272 | 153 | 196 |
| | 75th Percentile | 363 | 361 | 438 | 310 | 273 | 357 | 163 | 266 |
| Number of Stops per 1,000 mi | Mean | 3.58 | 3.64 | 3.18 | 3.99 | 4.03 | 3.60 | 6.41 | 5.18 |
| | Standard Deviation | 1.10 | 1.12 | 0.81 | 1.26 | 0.76 | 1.12 | 1.38 | 1.89 |
| | 25th Percentile | 2.75 | 2.77 | 2.28 | 3.22 | 3.66 | 2.80 | 6.12 | 3.76 |
| | Median | 3.65 | 3.60 | 3.11 | 3.75 | 4.03 | 3.67 | 6.54 | 5.10 |
| | 75th Percentile | 4.37 | 4.63 | 3.99 | 4.99 | 4.46 | 4.53 | 7.24 | 5.79 |

| (a) Section 5.1 Inclusion of Capital Costs | | | | | | | | |
|---|---|---|---|---|---|---|---|---|
| Commodity Group Name | | Agricultural & Foods | Chemical & Petroleum | Coal | Forest Products | Intermodal | Metals & Ores | Motor Vehicles | Others |
| Locomotive cost % | Mean | 45.2 | 47.0 | 49.7 | 40.2 | 11.6 | 46.9 | 23.3 | 26.2 |
| | Standard Deviation | 6.3 | 7.4 | 6.5 | 6.4 | 2.3 | 6.7 | 4.5 | 7.9 |
| | 25th Percentile | 39.8 | 41.6 | 43.1 | 36.2 | 10.6 | 41.6 | 22.1 | 22.6 |
| | Median | 46.3 | 47.5 | 51.5 | 38.1 | 11.3 | 47.5 | 23.2 | 25.0 |
| | 75th Percentile | 49.4 | 51.3 | 54.3 | 45.6 | 12.6 | 51.0 | 24.6 | 27.8 |
| Battery cost % | Mean | 14.7 | 14.3 | 15.4 | 15.6 | 12.1 | 14.5 | 12.6 | 15.3 |
| | Standard Deviation | 3.5 | 3.2 | 3.6 | 3.2 | 1.4 | 3.4 | 1.6 | 2.2 |
| | 25th Percentile | 11.9 | 11.8 | 12.8 | 11.7 | 11.0 | 11.8 | 11.4 | 13.5 |
| | Median | 12.6 | 12.7 | 13.4 | 16.9 | 12.0 | 12.8 | 12.1 | 15.3 |
| | 75th Percentile | 19.0 | 17.5 | 20.2 | 18.4 | 13.1 | 18.5 | 13.6 | 17.0 |
| Charging cost % | Mean | 28.9 | 27.2 | 25.5 | 32.5 | 34.0 | 26.5 | 40.5 | 34.6 |
| | Standard Deviation | 5.3 | 5.8 | 4.4 | 5.8 | 4.5 | 5.3 | 4.0 | 8.4 |
| | 25th Percentile | 26.4 | 24.3 | 23.7 | 28.9 | 31.7 | 23.7 | 39.7 | 32.5 |
| | Median | 28.9 | 27.5 | 24.9 | 32.0 | 34.2 | 27.1 | 41.3 | 34.6 |
| | 75th Percentile | 32.5 | 29.7 | 28.8 | 37.0 | 36.8 | 29.3 | 42.1 | 38.4 |
| Delay cost % | Mean | 11.3 | 11.5 | 9.4 | 11.7 | 42.2 | 12.1 | 23.5 | 23.9 |
| | Standard Deviation | 3.8 | 3.7 | 2.8 | 4.9 | 6.6 | 4.6 | 4.6 | 7.6 |
| | 25th Percentile | 9.2 | 9.4 | 8.1 | 7.9 | 37.9 | 9.0 | 21.2 | 18.7 |

| (a) Section 5.1 Inclusion of Capital Costs | | | | | | | | |
|---|---|---|---|---|---|---|---|---|
| **Commodity Group Name** | | **Agricultural & Foods** | **Chemical & Petroleum** | **Coal** | **Forest Products** | **Intermodal** | **Metals & Ores** | **Motor Vehicles** | **Others** |
| | **Median** | 10.3 | 11.2 | 8.5 | 10.7 | 42.3 | 11.6 | 23.7 | 24.7 |
| | **75th Percentile** | 12.8 | 12.6 | 10.8 | 13.1 | 46.4 | 12.7 | 24.9 | 26.2 |

| (b) Section 5.4 Exclusion of Capital Costs | | | | | | | | |
|---|---|---|---|---|---|---|---|---|
| **Commodity Group Name** | | **Agricultural & Foods** | **Chemical & Petroleum** | **Coal** | **Forest Products** | **Intermodal** | **Metals & Ores** | **Motor Vehicles** | **Others** |
| **Number of Batteries per Locomotives** | **Mean** | 2.61 | 2.52 | 2.91 | 2.56 | 5.29 | 2.53 | 3.18 | 3.67 |
| | **Standard Deviation** | 0.64 | 0.62 | 0.72 | 0.63 | 0.78 | 0.64 | 0.54 | 0.81 |
| | **25th Percentile** | 2 | 2 | 2 | 2 | 5 | 2 | 3 | 3 |
| | **Median** | 3 | 2 | 3 | 2 | 5 | 2 | 3 | 4 |
| | **75th Percentile** | 3 | 3 | 3 | 3 | 6 | 3 | 3 | 4 |
| **Range (mi)** | **Mean** | 597 | 610 | 754 | 444 | 311 | 619 | 229 | 314 |
| | **Standard Deviation** | 194 | 214 | 183 | 163 | 53 | 241 | 41 | 94 |
| | **25th Percentile** | 494 | 466 | 663 | 334 | 280 | 450 | 207 | 259 |
| | **Median** | 549 | 570 | 729 | 399 | 310 | 563 | 229 | 294 |
| | **75th Percentile** | 601 | 624 | 796 | 455 | 331 | 637 | 245 | 355 |
| **Number of Stops per 1,000 mi** | **Mean** | 1.80 | 1.78 | 1.36 | 2.49 | 3.30 | 1.79 | 4.50 | 3.45 |
| | **Standard Deviation** | 0.41 | 0.43 | 0.17 | 0.67 | 0.53 | 0.49 | 0.80 | 1.01 |
| | **25th Percentile** | 1.66 | 1.60 | 1.26 | 2.20 | 3.02 | 1.57 | 4.08 | 2.82 |

| (b) Section 5.4 Exclusion of Capital Costs | | | | | | | | | |
|---|---|---|---|---|---|---|---|---|---|
| **Commodity Group Name** | | **Agricultural & Foods** | **Chemical & Petroleum** | **Coal** | **Forest Products** | **Intermodal** | **Metals & Ores** | **Motor Vehicles** | **Others** |
| | **Median** | 1.82 | 1.76 | 1.37 | 2.51 | 3.23 | 1.78 | 4.36 | 3.40 |
| | **75th Percentile** | 2.02 | 2.15 | 1.51 | 3.00 | 3.57 | 2.22 | 4.82 | 3.86 |
| **Charging cost %** | **Mean** | 84.4 | 83.6 | 87.2 | 82.8 | 53.2 | 82.7 | 73.0 | 70.3 |
| | **Standard Deviation** | 4.9 | 5.1 | 2.9 | 5.7 | 6.4 | 5.5 | 5.1 | 8.2 |
| | **25th Percentile** | 83.2 | 83.1 | 85.7 | 83.1 | 49.6 | 83.1 | 71.7 | 67.2 |
| | **Median** | 85.9 | 84.9 | 88.3 | 83.9 | 53.3 | 83.5 | 73.8 | 67.2 |
| | **75th Percentile** | 87.6 | 86.1 | 88.5 | 86.1 | 57.5 | 86.0 | 75.9 | 76.1 |
| **Delay cost %** | **Mean** | 15.6 | 16.4 | 12.8 | 17.2 | 46.8 | 17.3 | 27.0 | 29.7 |
| | **Standard Deviation** | 4.9 | 5.1 | 2.9 | 5.7 | 6.4 | 5.5 | 5.1 | 8.2 |
| | **25th Percentile** | 12.4 | 13.9 | 11.5 | 13.9 | 42.5 | 14.0 | 24.1 | 23.9 |
| | **Median** | 14.1 | 15.1 | 11.7 | 16.1 | 46.7 | 16.5 | 26.2 | 32.8 |
| | **75th Percentile** | 16.8 | 16.9 | 14.3 | 16.9 | 50.4 | 16.9 | 28.3 | 32.8 |

10